\documentclass[amsmath,a4paper,twocolumn,pre]{revtex4}
\usepackage{graphicx}
\usepackage{amsmath}
\usepackage[english]{babel}
\usepackage{color}
\usepackage{braket}
\DeclareMathOperator{\Tr}{Tr}
\newcommand{\angstrom}{\mbox{\normalfont\AA}}

\begin{document}
\title{Dynamics of coherence, localization and excitation transfer in disordered nanorings}

%\title{Disorder effects on the coherence dynamics in light-harvesting complexes in purple bacteria}

\author{Alejandro D. Somoza$^{1}$, Ke-Wei Sun$^{2}$, Rafael A. Molina$^{3}$, and Yang Zhao$^1$\footnote{Electronic address:~\url{YZhao@ntu.edu.sg}}}

\affiliation{\it
$^{1}$Division of Materials Science,
Nanyang Technological University, 50 Nanyang Avenue, Singapore 639798\\
$^2$School of Science, Hangzhou Dianzi University, Hangzhou 310018, China\\
$^3$Instituto de Estructura de la Materia, IEM-CSIC, Serrano 123, Madrid 28006, Spain
}
\date{\today}

\begin{abstract}
Self-assembled supramolecular aggregates are excellent candidates for the design of efficient excitation transport devices. Both artificially prepared and natural photosynthetic aggregates in plants and bacteria present an important degree of disorder that is supposed to hinder excitation transport. { Besides, molecular excitations couple to nuclear motion affecting excitation transport in a variety of ways. We present an exhaustive study of exciton dynamics in disordered nanorings with long-range interactions under the influence of a phonon bath and take the LH2 system of purple bacteria as a model. Nuclear motion is explicitly taken into account by employing the Davydov ansatz description of the polaron and quantum dynamics are obtained using a time-dependent variational method. We reveal an optimal exciton-phonon coupling that suppresses disorder-induced localization and facilitate excitation de-trapping. This excitation transfer enhancement, mediated by environmental phonons,  is attributed to energy relaxation toward extended, low-energy excitons provided by the precise LH2 geometry with anti-parallel dipoles and long-range interactions. An analysis of localization and spectral statistics is followed by dynamical measures of coherence and localization, transfer efficiency and superradiance. Linear absorption, 2D photon-echo spectra and diffusion measures of the exciton are examined to monitor the diffusive behavior as a function of the strengths of disorder and exciton-phonon coupling.}
\end{abstract}

\maketitle

\section{Introduction}

Nanorings and other one-dimensional molecular aggregates are of great interest in the design of sensors and photovoltaic devices. The use of nano-structured rings as a probe for precise thermometry of quantum samples have been proposed \cite{Guo2015}, with applications to various branches of materials science, biology and physics. Engineered quantum rings may also serve as highly sensitive superabsorbing photonic sensors \cite{Higgins2014}. Artificial assemblies of chromophoric rings have also been considered, although many attempts resulted in dynamics that can be explained by incoherent hopping \cite{Cho2003}. Yong \textit{et al.}~\cite{Yong2015} have recently successfully demonstrated exciton delocalization in synthetic nanorings using porphyrin compounds.

Ring-shaped bacteriochlorophyll (BChl) aggregates are naturally found in the intracellular photochromatic membrane of purple bacteria showing hexagonal packing. Photons are captured by ring shaped light-harvesting 2 (LH2)  complexes in the form of an exciton that migrates to the reaction center contained in larger ring structures denominated light-harvesting 1 (LH1)\cite{Jang2001,Hu2002,Sundstrom1999}. Transfer of the exciton to the reaction center, where charge separation takes place proceeds with remarkable efficiency \cite{Scholes2011}. These photosynthetic systems receive the name of pigment-protein complexes because they are constituted by photosensitive chromophores (BChl-a in the case of purple bacteria) embedded in a protein environment that is strongly coupled to the electronic excitations of the pigment aggregate. The peripheral antenna complex
LH2 is a circular aggregate built from $\alpha\beta$-heterodimers. Depending on the species it presents $C_8$ or $C_9$ symmetry. Each unit has a pair of $\alpha$ and $\beta$ apoproteins, a carotenoid and three bacteriochlorophylls-a (BChls a) molecules \cite{McDermott1995,Koepke1996,Zhao2004a}. The BChl-a molecules in the LH2 form two rings, B800 and B850, named according to the wavelenghths of their absorption peaks at 800 nm and 850 nm. The chromophores of the B850 band are more tightly coupled to each other giving rise to delocalized excitonic states, while the chromophores in the B800 band are simply modeled by excitation hopping from one site to another \cite{Scholes2011}. In this work we will focus on exciton dynamics in the B850 band on a disordered energy landscape under environmental phonons.

Self-assembled artificial ring structures \cite{Yong2015} and natural photosynthetic aggregates often present a non-negligible degree of structural deformation at each pigment. Recent spectroscopic studies of the LH2 have shown that relaxation processes are highly dependent on the realization of disorder \cite{Novoderezhkin2011b}. On the top of this inhomogeneous broadening or static disorder, dynamic fluctuations such as solvent effects, nuclear motion and pigment-protein interactions are coupled to the electronic molecular excitations leading to homogeneous broadening of the spectral features. The transition dipole moments of the pigments in the B850 band are aligned in a tangential, in-plane configuration, with adjacent dipoles running in opposing directions (Figure \ref{fig:LH2headtotail}). This configuration induces an optical red-shift of the absorption band with respect to the monomer absorption peak. The electronic coupling between adjacent dipoles are positive, while the long-range interactions have alternating signs. Cyclic aggregates such as the LH2 present a sharp distribution of oscillator strengths where most of the optical
activity is held by a few superradiant, bright states \cite{Meier1997}. Superradiance relies on the cooperative
effect of coupled chromophores, where a single excitation can be shared among an extended patch of pigments,
leading to an enhanced radiative decay \cite{Dicke1954}. It is thus a quantum phenomenon, fully dependent on
the coherence between molecular excited states.  The optical activity under static disorder redistributes, lowering their cooperative nature. In the clean, non-disordered crystallographic structure of the LH2, all the oscillator strength lies within the degenerate states at the bottom of the exciton band $k=\pm 1$, immediately above the lowest energy state $k=0$, which is dark \cite{Note1}. In disordered rings, degeneracies are lifted and the oscillator strength redistributes, as revealed by single molecule spectroscopy for the lost of degeneracy of the $k=\pm 1$ states\cite{VanOijen1999}. Jang and Silbe \cite{Jang2003a,Jang2003} have extensively studied the splitting energy dependence of the bright states, concluding that both diagonal and off-diagonal disorder may be required to reproduce experimental values of the energy splitting. Additionally, static disorder induces localization of the wave functions and hinders diffusion in a tight binding model, a mechanism first described by Anderson \cite{Anderson1956,Evers2008}. In contrast, delocalization originated in long-range dipole-dipole interactions is more robust against disorder \cite{Rodriguez2003}. Unsurprisingly, long-range electronic coupling is found to enhance exciton transfer in comparison to a tight-binding model \cite{Ye2012}.

In the late 1990s, Zhao {\it et al.}~calculated distributions of cooperative radiative decay rates in photosynthetic aggregates in the presence of static diagonal disorder, which can be directly observed in single-molecule spectroscopy \cite{YangZhao1999JPCB}. Comparisons were also made with excitonic localization due to self-trapping.
{Str\"{u}mpfer and Schulten \cite{Strumpfer2009,StrumpferHEOM}} recently studied the dynamics of coherence in the B850 band by means of a Hierarchy of Equations of Motion (HEOM),
finding an steady-state coherence size of 11.2 pigments at 300 K for an initially localized state in an 18-pigment LH2, although their model does not address static disorder or the combined effect of static disorder and exciton-phonon coupling on excitonic localization \cite{Emin1994}.
Freiberg \textit{et al.} \cite{Freiberg2009} performed spectroscopic studies
concluding that the lowest-energy optical excitations in the LH1 and LH2 antenna complex are dominated by polaronic states. Static disorder may also be responsible for the emergence
of small-polaron states  in molecular aggregates \cite{Emin1994}. The interplay of vibronic interactions, energetic disorder and thermal activation requires careful evaluation
in order to establish a consensus on excitation energy transfer in photosynthetic systems. Recently, non-linear spectra of the LH2 has been obtained through atomistic modeling
and a combination of quantum-classical methods \cite{VanderVegte2015}. Polaronic effects in the LH2 at finite temperatures have been recently analyzed \cite{Chorosajev2016},
but the authors did not model excitation transfer including dissipative terms and their analysis of static disorder is scarce. The single Davydov $\rm D_2$ trial wave
function employed in that work is the simplest form of soliton-like trial states and only yields reliable results in the strong exciton-phonon coupling regime \cite{Forner1993}.
A recent analysis using the multiple-$\rm D_2$ trial state \cite{Zhou2015} reveals that when the electronic coupling between chromophores is strong,
a superposition of up to 32 $\rm D_2$ {\it ans\"{a}tze }
is required so that the accuracy of the trial state remains close to that of a single Davydov $\rm D_1$ {\it ansatz}, which is the variational state utilized in the present work.
We employ the Frenkel-Dirac time dependent variational method with the Davydov $\rm D_1$ {\it ansatz} and study the effects of static disorder on the coherence and excitonic dynamics of the
B850 ring {with vibronic coupling to intra-molecular modes}. Due to the computational overload of statistically meaningful disordered ensembles, we exploit the parallel architecture of
Graphic Processor Units (GPU) and deploy our dynamics simulations on this platform. {Further details on numerical considerations can be found in Appendix E.}

The rest of the paper is organized as follows. In Section II we will explain the model under consideration, the methodology employed and a justification
for the choice of system parameters. In Section III we will first analyze the disorder-induced localization in the absence of the phonon environment, which is followed by
results on exciton dynamics for different strengths of disorder and exciton-phonon coupling.
In addition, we will study a few measures of delocalization, and quantify the dephasing effects of the bath. Next, we will verify that exciton-phonon interactions
in disordered nanorings do enhance excitation transfer. Finally, the absorption spectra will be derived and the mean squared displacement of exciton diffusion will also be discussed. Conclusions are drawn in Section IV.

\section{Theory}

In this work we are going to model exciton dynamics in the B850 band of the LH2 pigment-protein complex from \textit{Rs.~molischianum}, whose crystal structure was resolved using X-ray diffraction by Koepke \textit{et al.} \cite{Koepke1996}. This aggregate presents $C_8$ symmetry and it is therefore constituted by 16 pigments. Details about the structural parameters can be found in Figure \ref{fig:LH2headtotail}.
We adopt a one-dimensional Holstein molecular crystal model \cite{HolsteinPolaronPart1,HolsteinPolaronPart2} with periodic boundary conditions in order to describe exciton dynamics in the LH2 embedded in a phonon bath. We have successfully utilized this formalism in previous studies for the analysis of polaron dynamics in purple bacteria \cite{Sun2010,Ye2012,Sun2014b}. Polaron states have been recently proposed to play a crucial role in the dynamics and directionality of energy transfer in the LH2 \cite{Freiberg2009,Ferretti2016}. The Holstein Hamiltonian may be written in the following form:
\begin{equation}\label{eq:Hamiltonian}
\hat{H}=\hat{H}_{\rm{ex}}+\hat{H}_{\rm{ph}}+\hat{H}_{\rm{ex-ph}}
\end{equation}
where $\hat{H}_{\rm{ex}}$ is the Frenkel-exciton Hamiltonian, $\hat{H}_{\rm{ph}}$ denotes the phonon Hamiltonian, and $\hat{H}_{\rm{ex-ph}}$ represents the exciton-phonon interactions. Each term of the Holstein Hamiltonian can be written as
{\begin{equation}\label{eq:Hex}
\hat{H}_{\rm{ex}}=\sum_{\rm nm} K_{\rm nm}\hat{a}_{\rm n}^{\dag}\hat{a}_{\rm m}
\end{equation}}
\begin{equation}\label{eq:Hph}
\hat{H}_{\rm{ph}}=\sum_{\rm q}\omega_{\rm q}\hat{b}_{\rm q}^{\dag}\hat{b}_{\rm q}
\end{equation}
\begin{equation}\label{eq:Hexph}
\hat{H}_{\rm{ex-ph}}=-\frac{1}{\sqrt{N}}\sum_{\rm nq}g_{\rm q}\omega_{\rm q}\hat{a}_{\rm n}^{\dag}\hat{a}_{\rm n}\left(e^{i \rm qn}\hat{b}_{\rm q}+H.c.\right)
\end{equation}
where the Planck's constant $\hbar$ has been set to one, $\hat{a}_{\rm n}^{\dag}$ $\left(\hat{a}_{\rm n}\right)$ is the creation (annihilation) operator of a molecular excitation at the n-th site, $b_{\rm q}^\dag$ ($b_{\rm q}$) creates (destroys) a phonon with momentum $\rm q$ and frequency $\omega_{\rm q}$, $g_{\rm q}$ determines the exciton's  coupling to phonons with momentum $\rm q$, and {$K_{\rm nm}$} is the Frenkel exciton Hamiltonian for the ring:
{\begin{equation}\label{eq:Jnm}
K_{\rm nm}=
\left(
\begin{array}{ccccccc}
\epsilon_{1} & J_{1} & W_{1,3} & \cdot & \cdot  & \cdot & J_{2}  \\
J_{1}& \epsilon_{2} & J_{2}  & \cdot & \cdot & \cdot & W_{2,16}  \\
W_{3,1} & J_{2} & \epsilon_{3} & \cdot & \cdot  & \cdot & \cdot  \\
\cdot & \cdot  & \cdot  & \cdot  & \cdot  & \cdot  & \cdot  \\
\cdot  & \cdot  & \cdot  & \cdot &\epsilon_{14} & J_{2} & W_{14,16} \\
\cdot  & \cdot  & \cdot  & \cdot& J_{2} &\epsilon_{15}  & J_{1} \\
J_{2}  & \cdot  & \cdot  & \cdot & W_{16,14} & J_{1} &\epsilon_{16}    \\
\end{array}
\right)
\end{equation}}
For the site energies $\epsilon_{i}$ of the Hamiltonian, we neglect the small differences between the chemical environments of the $\alpha$ and $\beta$ pigments and consider equal excitation energies for every pigment unless otherwise specified:
\begin{equation}\label{6}
\epsilon_{i}=\epsilon_{i}^{\prime}+\delta\epsilon_{i} \quad i\in \left[1,N=16\right]
\end{equation}
where we set $\epsilon_{i}^{\prime}=0$ for simplicity and $\delta\epsilon_{i}$ is the diagonal disorder term. Dimerization in the electronic coupling between nearest neighbors has been shown to affect the excitation transfer dramatically \cite{XuDazhiDimerization}. We have extracted these parameters from a semiempirical INDO/S analysis \cite{Zhao2004a}:
\begin{equation}\label{7}
J_{i}=J_{i}^{\prime}+\delta{}J_{i} \quad i\in \left[1,2\right]
\end{equation}
with the intra-dimer coupling constant $J_{1}^{\prime}=594$ cm$^{-1}$ and the inter-dimer coupling constant $J_{2}^{\prime}=491$ cm$^{-1}$. The term $\delta J_{i}$  accounts for the off-diagonal disorder. In accordance with an extensive analysis of static disorder in the LH2 by Jang {\it et al.}~\citep{Jang2001} and single-molecule experiments \cite{VanOijen1999,Hofmann2004} we will employ Gaussian distributions in order to model both sources of disorder. It is believed that both types of disorder are simultaneously present in the LH2, with an estimated $\sigma$ { ranging from $100$ to $200$ cm$^{-1}$} \cite{Wu1997,VanOijen1999,Jang2001}. We have also corroborated that the employed Gaussian distributions give rise to the same apparent behaviour of the variables under consideration by using geometrical distortion in the structural parameters. Thus, both $\delta\epsilon_i$ and $\delta J_{i}$ are sampled from independent Gaussian distributions:
\begin{equation}\label{8}
P(\delta{}\epsilon_{i},\delta{}J_{j})=\frac{1}{\sqrt{2\pi \sigma_{{E}}^2}}\frac{1}{\sqrt{2\pi \sigma_{{J}}^2}}\exp\bigg(\frac{\delta{}\epsilon_{i}^{2}}{2\sigma_{E}^{2}}+\frac{\delta{}J_{j}^{2}}{2\sigma_{J}^{2}}\bigg)
\end{equation}
\noindent where the standard deviations $\sigma_{\rm E}$, $\sigma_{\rm J}$ correspond to diagonal and off-diagonal disorder respectively. { As a word of caution, we remind the reader that ensembles of diagonal and off-diagonal disorder with equal standard deviations ($\sigma_{\rm E}=\sigma_{\rm J})$ are not strictly comparable to each other: for large values of $\sigma_{\rm J}$, off-diagonal disorder will effectively decouple the ring at random positions, acutely affecting localisation and coherence. Besides, the sign of some coupling elements may be reversed. This should be kept in mind when interpreting the following results.} The long-range matrix elements $W_{\rm ij}$ are assumed to arise from dipole-dipole interactions between non-nearest neighbors chromophores \cite{Zhao,Ye2012}:
\begin{equation}\label{eq:dipoledipole}
W_{\rm ij}=C \left[\frac{\hat{\textbf{d}}_{\rm i}\cdot\hat{\textbf{d}}_{\rm j}}{|\textbf{r}_{\rm ij}|^{3}}-\frac{3(\hat{\textbf{d}}_{\rm i}\cdot\textbf{r}_{\rm ij})(\hat{\textbf{d}}_{\rm i}\cdot\textbf{r}_{\rm ij})}{|\textbf{r}_{\rm ij}|^{5}}\right]
\end{equation}
where $C$ is the proportionality constant $C=640725{\angstrom}^3\rm{cm}^{-1}$ \cite{Zhao2004a}, $\textbf{r}_{\rm ij}$ is the distance vector between the $\rm {i}^{th}$ and $\rm{j}^{th}$ monomers, and $\hat{\textbf{d}}_{\rm i}$ is a unit vector characterizing the transition dipole moment of the $\rm{i}^{th}$ chromophore.

Regarding the phonon environment, we are going to focus on a resonant intramolecular $\omega_0=1670$ cm$^{-1}\sim20$ fs vibration { which linearly couples to the electronic excited states of the LH2's chromophores. This can be clearly seen in the structure of the interaction Hamiltonian (Eq.~\ref{eq:Hexph}). This form of interaction results in dynamic fluctuations of the optical transitions of each chromophore, as shown in a combined quantum chemistry, molecular dynamics study by Damjanovi{\'{c}} \textit{et al.} \cite{Damjanovic2002}}. A recent atomistic calculation of the spectral density in the LH2 also predicts resonant peaks around $\omega_0$ \cite{Olbrich2010}. It is thus reasonable to expect a certain dispersion of this mode in the aggregate. For this reason, we adopt a linear dispersion with a tunable phonon bandwidth $W\in\left[0,1\right]$ in accordance with previous studies of the LH2. \cite{Tanaka2003,Ye2012,Sun2014b}:
\begin{figure}[htb!]\label{fig:LH2headtotail}
  \includegraphics[scale=0.4]{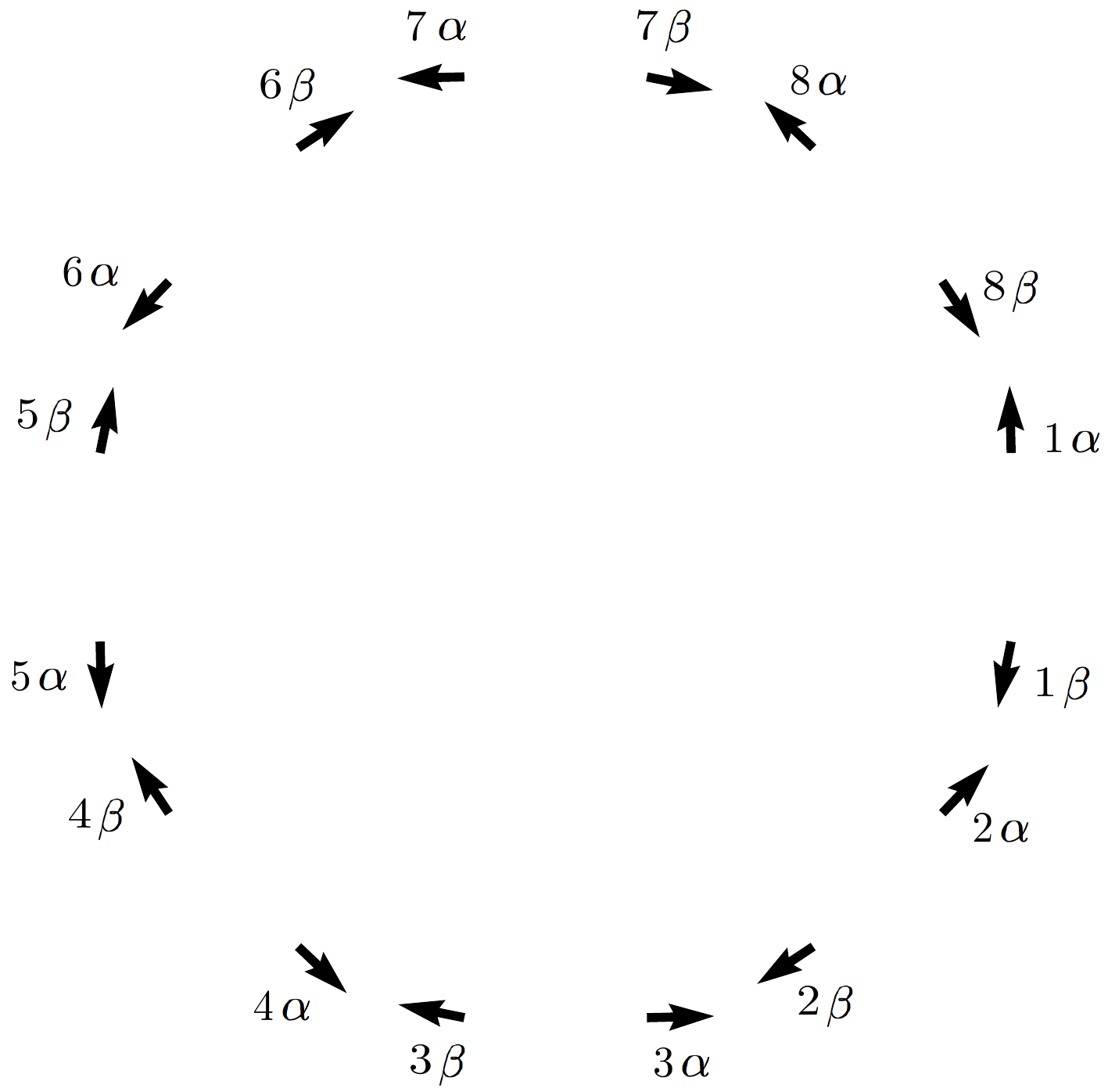}
  \caption{Schematic of the orientation of the $Q_y$ transition dipole moments in the B850 ring from \textit{Rs. molischianum} with $C_8$ symmetry. The intra-dimer and inter-dimer distances are 9.1 $\angstrom$ and 8.9 $\angstrom$ repsectively and the radius is 23 $\angstrom$. The angles between transition dipole moments are 167.5$^\circ$ (intra-dimer) and 147.5$^\circ$ (inter-dimer)  leading to a head-to-tail configuration. Nearest neighbors coupling constants are extracted by a semiempirical method \cite{Zhao2004a}: $J_{1}^{\prime}=594$ cm$^{-1}$ for intra-dimer pigments and $J_{2}^{\prime}=491$ cm$^{-1}$ for inter-dimer pigments. Long range interactions are calculated using dipole-dipole assuming dipole-dipole interactions (Eq.~\ref{eq:dipoledipole}).}
\end{figure}
\begin{equation}
\omega_{\rm q}=\omega_{0}+2W\omega_0\left(\frac{|q|}{\pi}-\frac{1}{2}\right)
\label{eq:dispersionrelationship}
\end{equation} %$q=2\pi{}n_{\rm q} /16(n_{\rm q}=-7,-6,\dots,7,8)$
\noindent where $q=\frac{2\pi}{N} \left( -\frac{N}{2}+(i+1)\right) \quad i=0,...,N-1$. Einstein phonons can be modelled by setting $W=0$. The structure of the exciton-phonon interaction is captured by the spectral density function:
{\begin{equation}\label{eq:spectraldensity}
J_{\rm nm}(\omega)\equiv\frac{1}{2\pi}\int_{-\infty}^{\infty}\langle\hat{V}_{\rm m}(t)\hat{V}_{\rm n}(0)\rangle_{\rm ph}\emph{e}^{i\omega{}t}dt
\end{equation}}
where %$\hat{V}_{n}$: denotes the Heisenberg representation of the exciton-phonon interaction at the $\rm n^{th}$ site,
\begin{equation}\label{eq:Vn}
\hat{V}_{\rm n}=\frac{1}{\sqrt{N}}\sum_{\rm q}g_{\rm q}\omega_{\rm q}(e^{i\rm qn}\hat{b}_{\rm q}+e^{-i\rm qn}\hat{b}_{\rm q}^{\dag})
\end{equation}
and $\langle\cdot\cdot\cdot\rangle_{\rm ph}$ is the thermal equilibrium for the
ground state. Substituting Eq.~(\ref{eq:Vn}) into Eq.~(\ref{eq:spectraldensity}):
{\begin{equation}
J_{\rm nm}(\omega)=\frac{1}{N}\sum_{\rm q}g_{\rm q}^{2}\omega_{\rm q}^{2}e^{i\rm q(m-n)}\delta(\omega-\omega_{\rm q})
\end{equation}}
Following Tanaka \cite{Tanaka2003} and neglecting spatial correlations \cite{Olbrich2010,VanderVegte2015}, we employ the following elliptic form of the spectral density:
{\begin{equation}
J_{00}(\omega)= \frac{2S\omega^{2}}{N\pi{}W^{2}}\sqrt{W^{2}-(\omega-\omega_{0})^{2}}
\end{equation}}
\noindent where the Huang-Rhys factor S determines the global coupling strength of the exciton manifold to the nuclear motion: $\frac{1}{N}\sum_{\rm q}g_{\rm q}^{2}\omega_{\rm q}=S\omega_{0}=\Lambda$, where $\Lambda$ is the reorganization energy.

As a means of simulating the polaron dynamics we made use of a family of time dependent wave functions  of the $D_1$ type belonging to the Davydov \textit{ansatz} family:
\begin{equation}\label{eq:D1}
\ket{D_1(t)}=\sum_{\rm n}\alpha_{\rm n}(t)\hat{a}_{\rm n}^{\dag}\hat{U}_{\rm n}^{\dag}(t)\ket{0}_{\rm ex} \ket{0}_{\rm ph}
\end{equation}
where $\alpha_{n}$ is the amplitude of an exciton in the $\rm n^{th}$ site of the ring and $U_{\rm {n}}^\dag$ is the displacement operator which creates a product of coherent states at the ${\rm n}$-th site:
\begin{equation}\label{eq:Un}
\hat{U}_{\rm n}^{\dag}(t)\equiv{}\exp{{\sum_{\rm q}[\lambda_{\rm nq}(t)\hat{b}_{\rm q}-H.c.]}}
\end{equation}
where $H.c$ stands for Hermitian conjugate.
We can extract the equations of motion for the variational parameters from the Dirac-Frenkel Lagrangian formalism, which is described in Appendix B:
\begin{eqnarray}
\frac{d}{dt}\bigg(\frac{\partial{}\mathcal{L}}{\partial{}\dot{\beta}_{\rm n}^{*}} \bigg)-\frac{\partial{}\mathcal{L}}{\partial{}\beta_{\rm n}^{*}}=0,~~~~\beta_{\rm n}\in\{\alpha_{\rm n},\lambda_{\rm nq}\} \end{eqnarray}
where the Lagrangian takes the following form:
\begin{eqnarray}
\mathcal{L}=\langle D_1(t)|\frac{i}{2}\frac{\overleftrightarrow{\partial}}{\partial{}t}-\hat{H}|D_1(t)\rangle
\end{eqnarray}
Unless otherwise specified we will start with a local excitation in the ring $\alpha_{\rm n}(0)=\delta_{\rm n,8}$ with the vibrational bath in the ground state $\lambda_{\rm nq}=0$. An initially localized excitation is plausible scenario for the case of excitation transfer from a surrounding carotenoid molecule, which are known to be efficient donors in the LH2 \cite{Zhang2000,Polivka2004}. The second assumption implies that our dynamics capture the zero temperature limit: the resonant modes that we are studying present frequencies which cannot be thermally activated ($1600$ cm$^{-1} >> 200$ cm$^{-1}$) and they will get populated upon exciton injection. In a realistic system, $S = 0.5$ \cite{Damjanovic2002} and $\sigma$ is in the range of $100 \sim 200$ cm$^{-1}$ \cite{Wu1997,VanOijen1999,Jang2001}.

\begin{figure}[htb!]\label{fig:iprbeta}
  \includegraphics[scale=0.5]{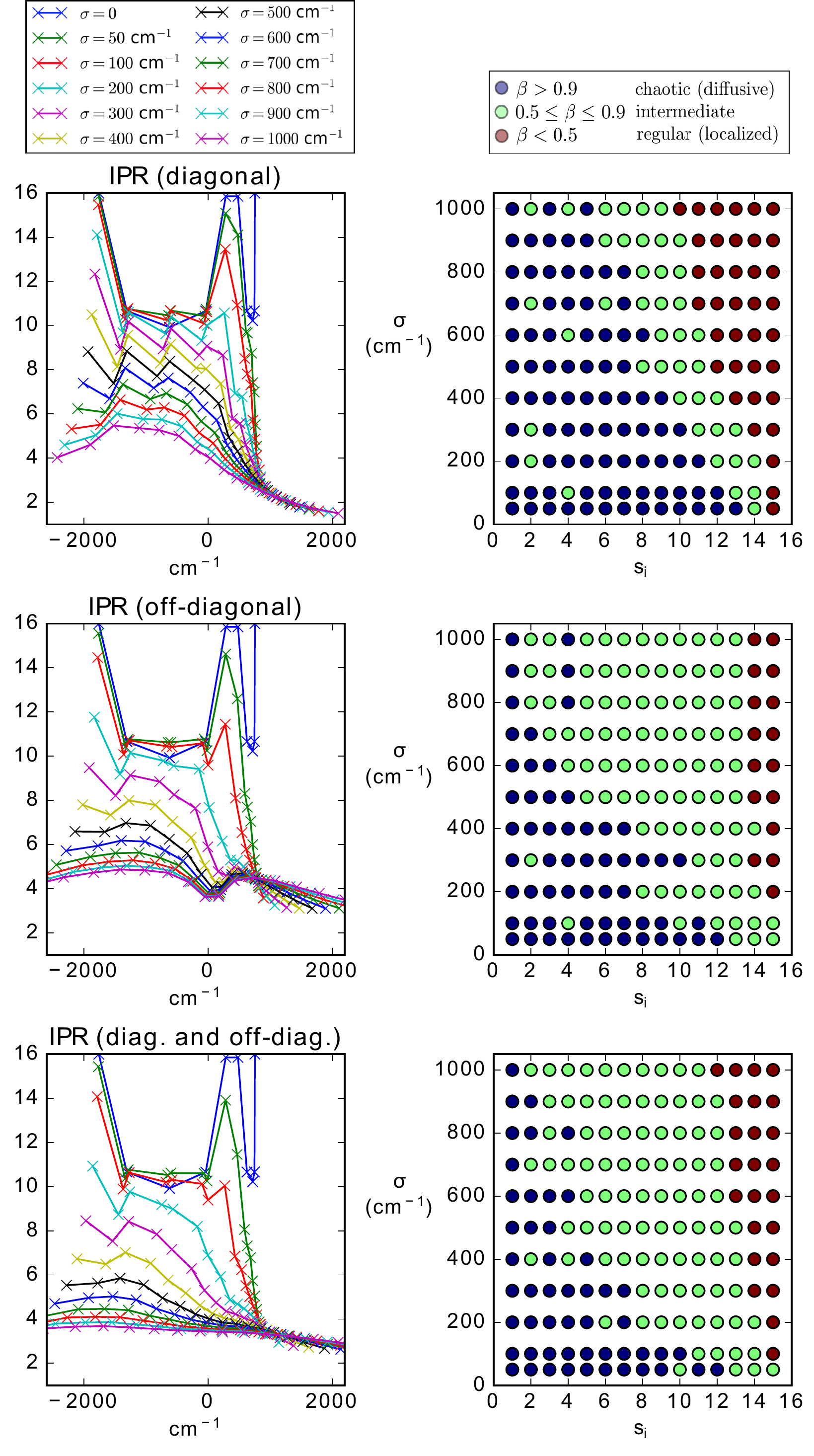}
  \caption{(Left column) IPR (Eq.~\ref{eq:IPR}) as a function of disorder strength and energy of the disordered excitonic spectrum. Growing downwards, the values of disorder are {$\sigma=0, 50, 100, 200, 300, 400, 500, 600, 700, 800, 900, 1000$ cm$^{-1}$}. The different panels correspond to diagonal disorder (top), off-diagonal disorder (center) and both diagonal and off-diagonal (bottom). (Right column) Repulsion parameter as defined in Eq.~\ref{eq:repulsion}. We have defined three different regions: $\beta > 0.9$, chaotic region or diffusive (blue dots); $0.5 \leq \beta \leq 0.9$, intermediate region (light green dots); $\beta<0.5$ localized or regular region (red dots). These results have been obtained after an ensemble average of 10000 realizations of disordered Hamiltonians. }
\end{figure}

\section{Results and Discussion}

\subsection{Localization and spectral statistics of a bare exciton}

{Let's investigate the impact of disorder in the absence of the phonon environment. The purpose of this section is to show that the systems under investigation present robust delocalised states, beneficial for transport, around the bottom of the band. These states are populated by energy relaxation via environmental phonons. {The long-range, parallel-antiparallel structure of the Hamiltonian is crucial for the validity of the following results. In contrast, a model with nearest-neighbour interactions (or parallel dipoles with long-range interactions) would present highly localised states at the bottom of the excitonic band, and phonon-mediated relaxation would severely hinder excitation transfer.}

A typical measure of localization is the inverse participation ratio (IPR) \cite{Bell70}:
\begin{equation}\label{eq:IPR}
 {\rm IPR}(E_{i})=\frac{1}{\sum_{\rm n} |\Psi_{\rm n}^{i}|^4}
\end{equation}
where $\Psi_{\rm n}^{i}$ is the amplitude of the excitonic eigenstate with energy $E_{i}$ on site $\rm n$. ${\rm IPR}=1$ if the state is localized on only one site, and ${\rm IPR}=N$ if the wave function is equally distributed among all sites in an $N$ site system. We will complement this information with an analysis of the repulsion parameter borrowed from Random Matrix Theory (RMT), a common tool for the analysis of wave localization and transport in disordered quantum systems \cite{Mehta_Book}. Specific scaling relations between wave function localization and the repulsion parameter in finite, disordered one-dimensional systems have been proven \cite{Izrailev89,Izrailev90,Casati93,Sorathia12}. { In particular, the repulsion parameter in one-dimensional systems is very sensitive to the ratio (localization length)/(system size) and therefore we can interpret the repulsion parameter as a complementary measure of wave function localization \cite{Sorathia12}. Further information regarding the extraction of the repulsion parameter can be found in Appendix A}.

We plot in Figure \ref{fig:iprbeta} the results of this disorder-induced localization analysis. In the left column we display the excitonic IPR for various disorder strengths for diagonal disorder (top panel), off-diagonal disorder (center panel) and both types of disorder {simultaneously present} (lower panel). Since we assume the same excitation energy for all sites in the ring, we have subtracted {this value from the eigenstates} for easier illustration. The most immediate { effect} of disorder is the {breakdown} of the doubly degenerate structure of the excitonic spectrum (with the exception of the ground state and the highest energy state in the pristine system, all eigenstates come in degenerate pairs). {Each pair of degenerate states branches} into two separate states which diverge { from each other} with increasing disorder. { The localization properties of the spectrum in the three cases show an asymmetric IPR with robust, delocalised states at low energies. For realistic values of disorder, where $\sigma$ ranges between 50 and 200 cm$^{-1}$, the IPR of the lowest excitonic states ranges between 10 and 12 sites. This is in stark contrast to a simple model with nearest-neighbour interactions, where the IPR of these states would be at least 4 sites lower, especially for strong disorder.} { In comparison with the diagonal case, off-diagonal disorder} widens slightly the range of the spectrum. This may be understood by considering the fact that the excitonic band gap is generally proportional to the inter-chromophoric coupling. {For large values of disorder,} there appears a dip around $0$ cm$^{-1}$ in the off-diagonal case which is flattened when diagonal disorder is also included. In all cases, localization is very sensitive to the exciton's energy, with the highest energy states having around half the IPR value of the low energy states.

As for the repulsion parameter $\beta$, {which is proportional to the localization length of wave functions \cite{Sorathia12}}, the results are shown in Figure \ref{fig:iprbeta} (right column) as a function of disorder strength and nearest neighbors spacing $s_i = (E_{i+1} -E_i)$. We have distinguished three different regions according to its value (Eq.~\ref{eq:repulsion}): $\beta > 0.9$ chaotic region or diffusive (blue dots), $0.5 \leq \beta \leq 0.9$ intermediate region (light green dots) and $\beta<0.5$ localized or regular region (red dots). {We observe how extended states at low energies, corresponding to a diffusive regime survive even for large values of disorder in the three cases of disorder.} Localization affects more strongly high energy states, which rapidly become localized (red dots). The extent of the localized region is larger in the diagonal disorder case. In the case of off-diagonal disorder, the localized region gets reduced but the intermediate region also increases at the expense of reducing the extent of the diffusive, delocalized region. When both diagonal and off-diagonal disorder, the impact of diagonal disorder in the high energy states gets significantly reduced.
\begin{figure*}[tbp]
  \centering
  \includegraphics[width=0.75\linewidth]{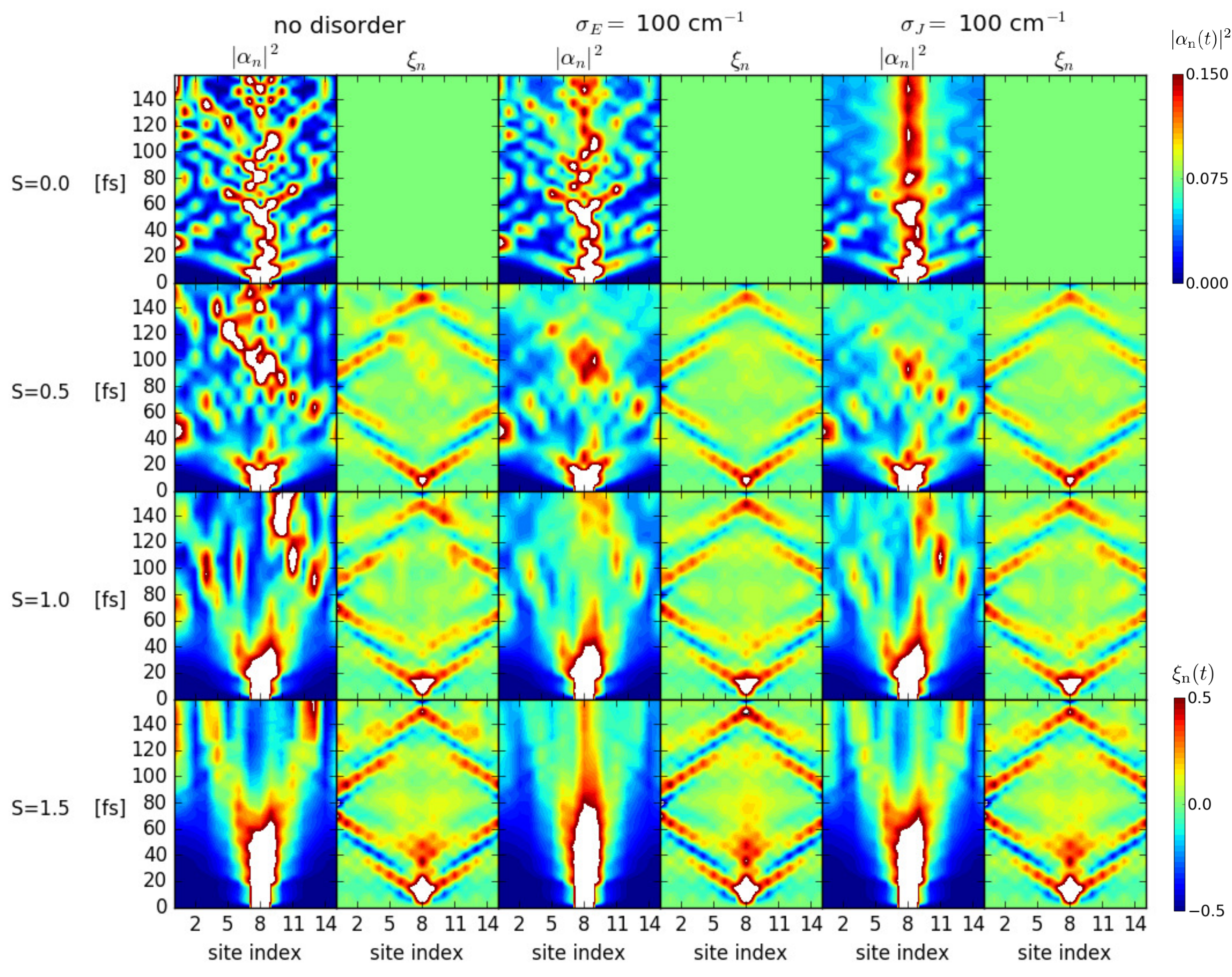}
  \caption{Real space dynamics of the exciton probability $|\alpha_{\rm n}(t)|^2$ and phonon displacement $\xi_{\rm n}(t)$. The left panels show the results without any static disorder. The center panels are computed with diagonal disorder ($\sigma_{\rm E}$=100 cm$^{-1}$) while the right panels are given with off-diagonal disorder ($\sigma_{\rm J}$=100 cm$^{-1}$). Different rows correspond to increasing exciton phonon coupling with corresponding Huang-Rhys factors of S = 0.0, S = 0.5, S = 1.0 and S = 1.5. The color scale for the exciton population is 0 for blue and red for 0.15, while regions with larger probability are white. The color code for the phonon displacements has been ranged in every case to $\xi_{\rm n}\in[-0.5,0.5]$ in order to stress the differences between weak and strong coupling cases. White regions appear whenever the displacement is outside this region. The results have been averaged over 1000 samples.
  }\label{fig:excphondynE100}
\end{figure*}
\subsection{Polaron dynamics}
Let us now analyze the dynamics of the exciton in a disordered energy landscape for various exciton-phonon coupling strengths. In analogy with the exciton population $|\alpha_{\rm n}(t)|^2$, we can study the phonon environment by introducing the phonon displacement $\xi_{\rm n}$ at site $\rm n$:
\begin{equation}
 \begin{split}
\xi_{\rm n}(t)&=\bra{D_1(t)}\frac{\hat{b}_{\rm n}^{\dag}+\hat{b}_{\rm n}}{2}\ket{D_1(t)} \\
&=\frac{1}{2\sqrt{N}}\sum_{\rm q}\sum_{\rm n^{\prime}}|\alpha_{\rm n^{\prime}}(t)|^2(e^{-i\rm qn}\lambda_{\rm n^{\prime}q}^{*}+e^{i\rm qn}\lambda_{\rm n^{\prime}q})
\end{split}
\end{equation}
where $\hat{b}_{\rm n}^{\dag}=N^{-1/2}\sum_{\rm q}e^{-i\rm qn}\hat{b}_{\rm q}^{\dag}$. The dynamics of exciton probability $|\alpha_{\rm n}(t)|^{2}$ and phonon displacement $\xi_{\rm n}(t)$ are shown in Figure \ref{fig:excphondynE100}, for a phonon bandwidth $W=0.5$ and fixed static disorder of $\sigma_{\rm E}=100$ cm$^{-1}$ (center panels) and $\sigma_{\rm J}=100$ cm$^{-1}$ (right panels). Calculations without static disorder are shown at left for comparison. Figure \ref{fig:excphondynE300} shows the same results for stronger static disorder with $\sigma_{\rm E}=300$ cm$^{-1}$ or $\sigma_{\rm J}=300$ cm$^{-1}$. {The main result of this section is clearly illustrated in Figure \ref{fig:excphondynE300} with $\sigma=300$ cm$^{-1}$, by comparing the first row ($S=0$, no phonons) with the second ($S=0.5$) and third row ($S=1.0$). When no phonons are present, the wave packet remains localized and it does not spread. When moderate coupling to vibrations is turned on, the excitation is allowed to diffuse. However, if the exciton-phonon coupling is too large, dynamic localization of the exciton sets in and the excitation is again localized. An optimal parameter regime with the Huang-Rhys factor $S$ around 0.5, corresponding to a reorganization energy of 835 cm$^{-1}$ is thus expected and will be confirmed in the forthcoming sections. This is precisely the estimated value of the exction-phonon coupling found in the LH2 \cite{Damjanovic2002}.} Further increasing the exciton-phonon coupling strength tends to slow down the initial diffusion of the exciton. In the absence of exciton phonon interactions ($S = 0$) we observe that population dynamics are somewhat symmetric with respect to the initial site. In contrast, when moderate exciton-phonon interaction takes place ($S = 0.5$), there appears a preferred direction where the exciton travels from right to left during the interval $30$ to $120$ fs. This behavior endures for weak static disorder $\sigma_{\rm E}, \sigma_{\rm J}=$ 100 cm$^{-1}$ (Figure \ref{fig:excphondynE100}) but disappears for stronger disorder $\sigma_{\rm E}, \sigma_{\rm J}=$ 300 cm$^{-1}$ (Figure \ref{fig:excphondynE300}). Regarding the phonon displacement, the dominant V-shaped feature of the displacement remains in the disordered case due to its own propagation pattern, which is determined by the employed linear dispersion relationship (\ref{eq:dispersionrelationship}). In particular, the phonon bandwidth parameter $W$ determines the group velocity of the phonon displacement which in the plots of Figures (\ref{fig:excphondynE100}) and (\ref{fig:excphondynE300}) translates into a steeper V-shape with increasing $W$. Nevertheless, on the top of the dispersion pattern there are phonon displacements which are due to energy exchange between the exciton and the phonon environment. {The differences between diagonal disorder and off-diagonal disorder are not very severe and overall they impact the exciton dynamics in similar manners. When disorder is strong (Fig.~\ref{fig:excphondynE300}), the off-diagonal case benefits from exciton-phonon coupling in any case. For diagonal disorder, an optimal value of $S$ is found.}
\begin{figure*}[tbp]
  \centering
  \includegraphics[width=0.75\linewidth]{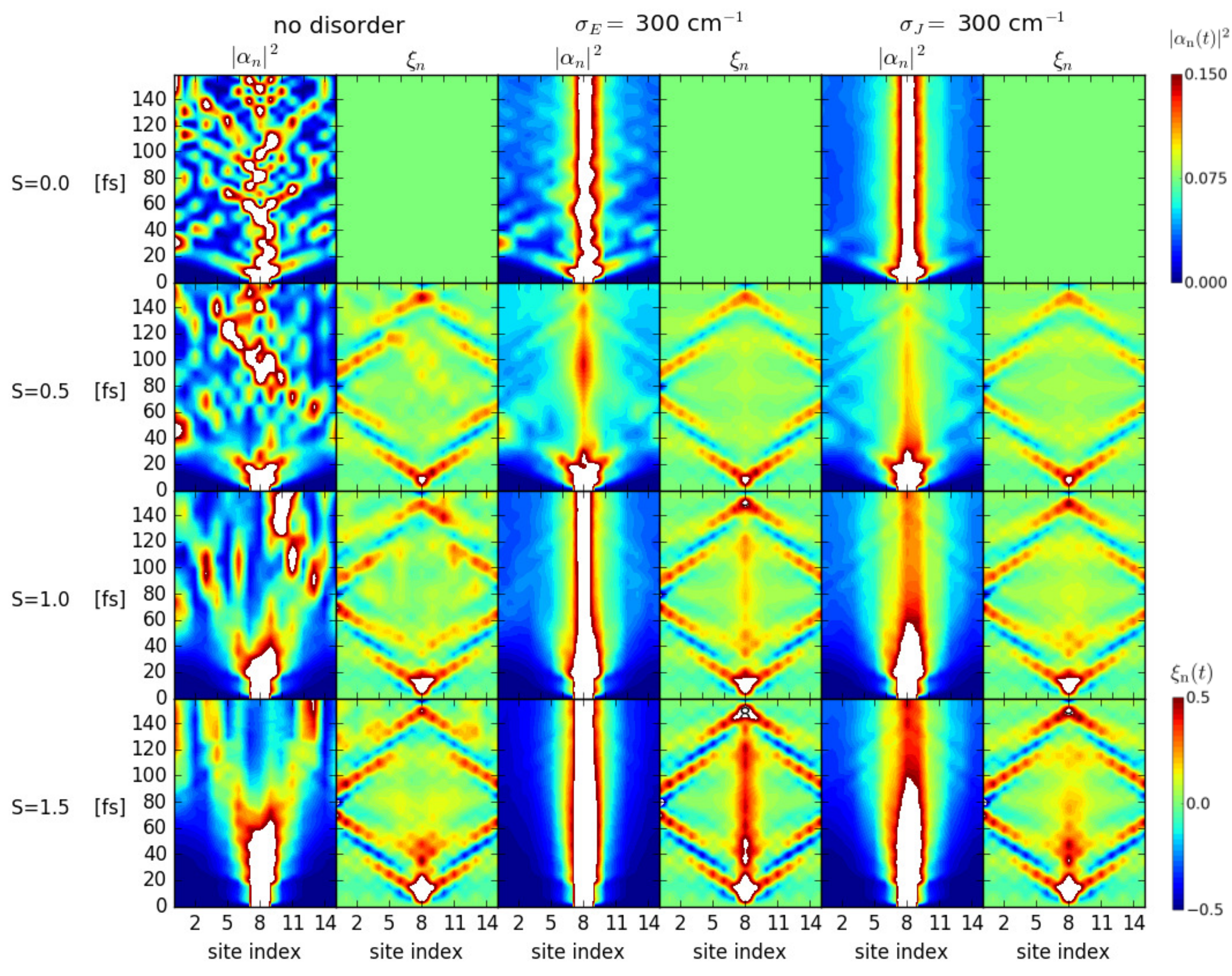}
  \caption{Real space dynamics of the exciton probability $|\alpha_{\rm n}(t)|^2$ and phonon displacement $\xi_{\rm n}(t)$. The left panels show the results without any static disorder. The center panels are computed with diagonal disorder ($\sigma_{\rm E}$=300 cm$^{-1}$) while the right panels are given with off-diagonal disorder ($\sigma_{\rm J}$=300 cm$^{-1}$). Different rows correspond to increasing exciton phonon coupling with corresponding Huang-Rhys factors of S = 0.0, S = 0.5, S = 1.0 and S = 1.5. {The same color scheme of Figure 3 has been employed.}
  }\label{fig:excphondynE300}
\end{figure*}
\begin{figure*}[htb!]
\includegraphics[scale=0.5]{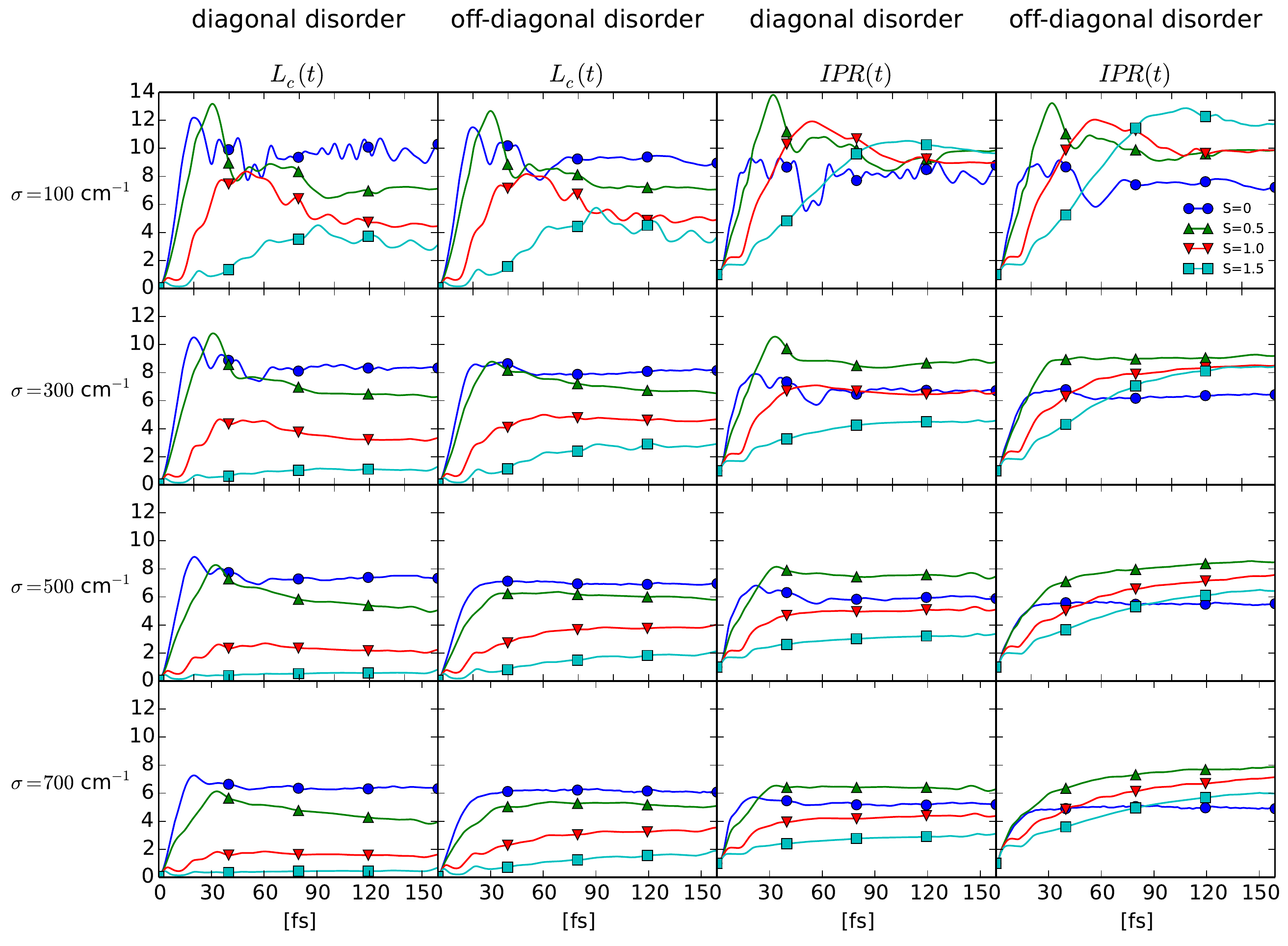}
\caption{ $L_c(t)$ for different Huang-Rhys factor (exciton-phonon coupling strength) $S$ with a bandwidth parameter of $W=0.5$. Left panels correspond to diagonal disorder and right panels to off-diagonal disorder. Different rows correspond to increasing standard deviation of the disorder $\sigma_{\rm{E}}$, $\sigma_{\rm{J}}$ with 100 cm$^{-1}$, 300 cm$^{-1}$, 500 cm$^{-1}$, 700 cm$^{-1}$. Circles correspond to the bare exciton model (no phonons); up-triangles, down-triangles and squares correspond to S = 0.5, S = 1.0 and S = 1.5 respectively.  The results have been averaged over 1000 samples.}
\label{fig:csizedynamics}
\end{figure*}
\begin{figure}[htb!]
  \centering
  \includegraphics[scale=0.5]{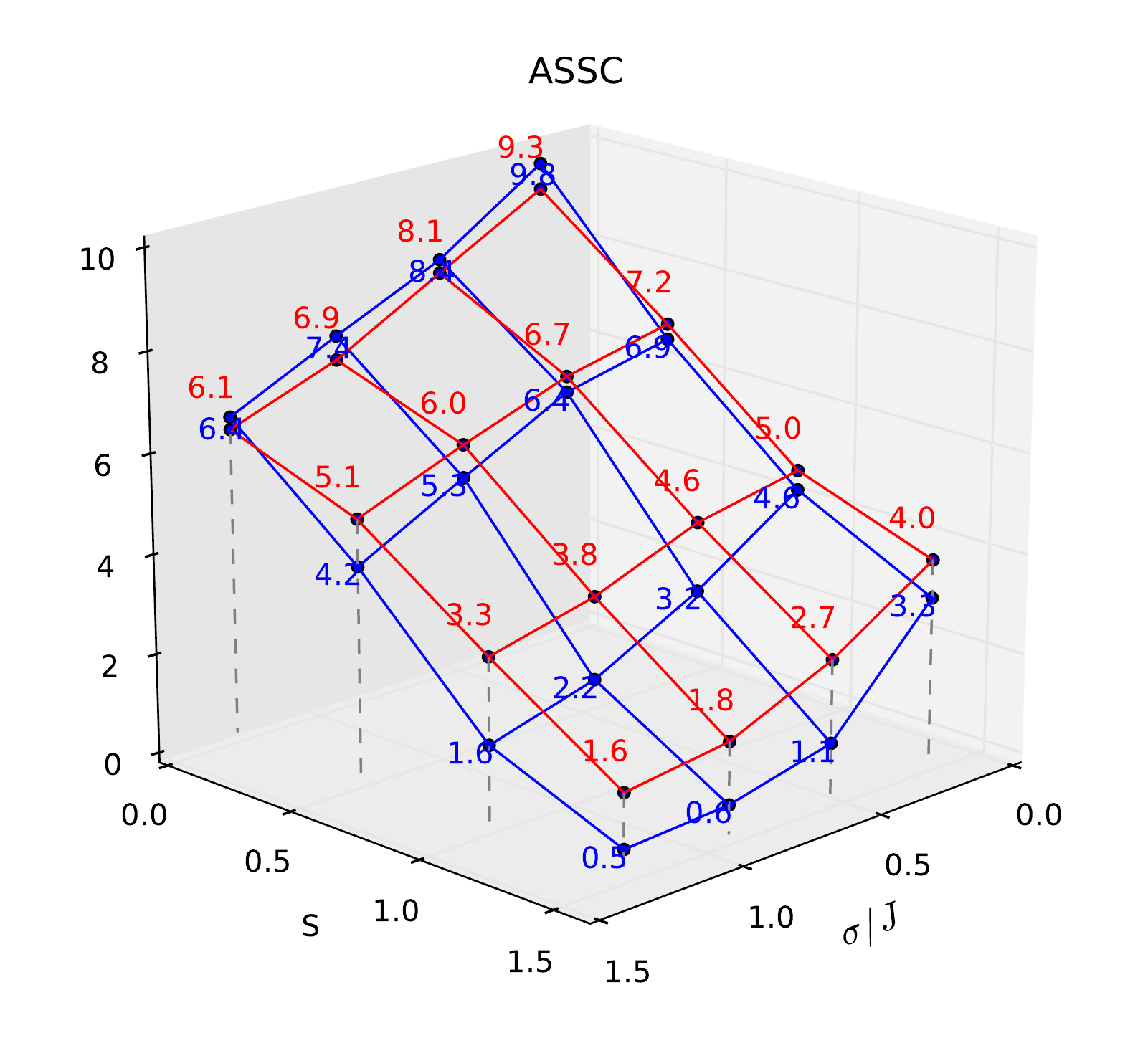}\\
  \caption{
Bird view plot of the average steady state coherence size ASSC. The results have been averaged over 1000 samples for a single ring system with $W=0.5$.
  }\label{fig:ASSC}
\end{figure}

\begin{figure}[htb!]
  \centering
  \includegraphics[scale=0.5]{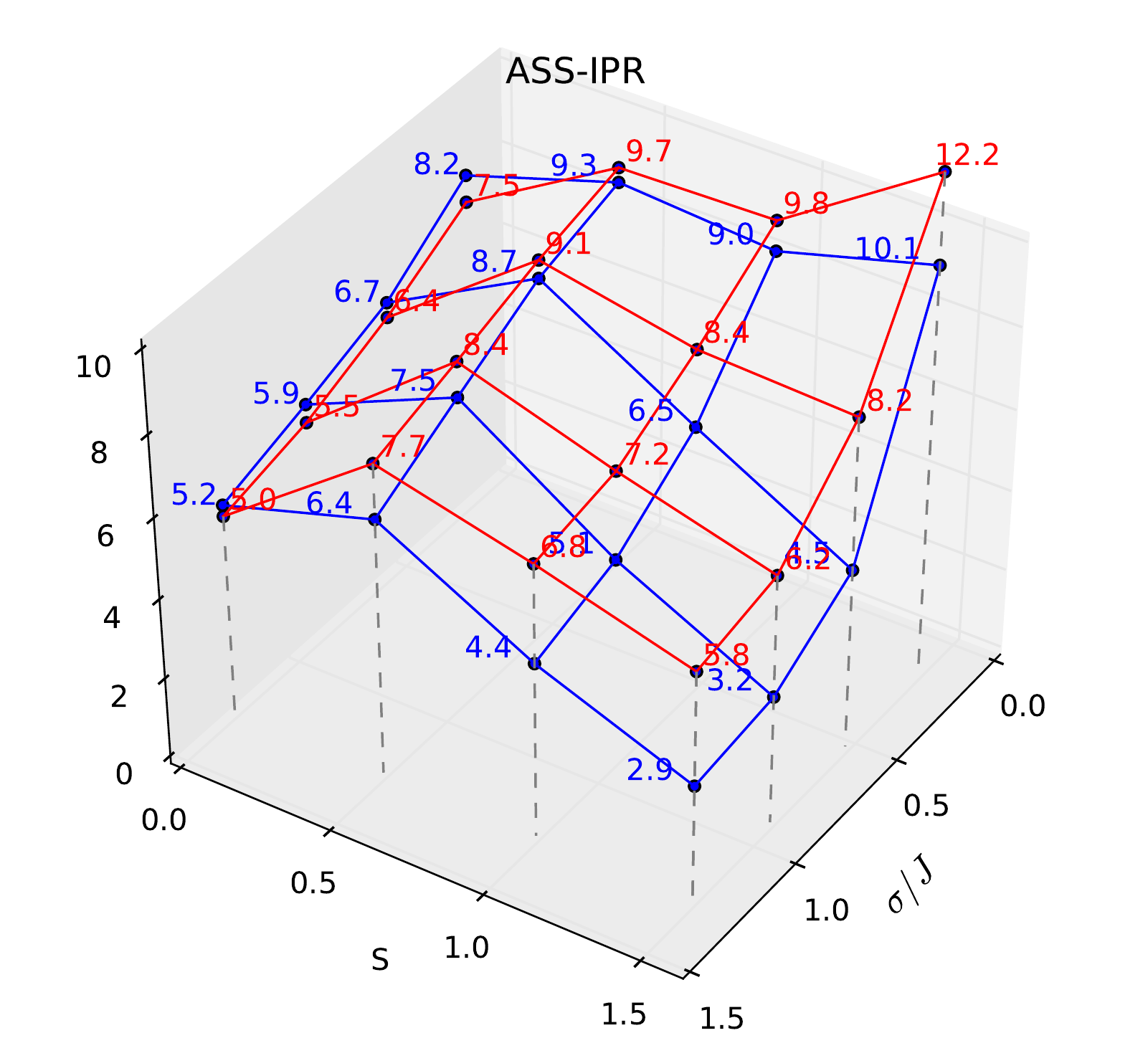}\\
  \caption{
  The 3D-plot of the average steady-state IPR. The results have been averaged over 1000 samples for a single ring system with $W=0.5$.
  }\label{fig:ASSIPR}
\end{figure}

\subsection{Coherence and Delocalization}

Whenever we employ the term \textit{coherence}, we will explicitly refer to coherence among the ${\rm N}$ localized molecular excited states $\lbrace\ket n\rbrace$ with ${n\in \left[1,{\rm N}\right]}$ that conform the site basis of the aggregate. This is an important remark since in spectroscopic experiments, on the contrary, coherences are probed among excitonic eigenstates. In the presence of static disorder, the excitonic wave functions suffer Anderson localization that can lead to the suppression of diffusion \cite{Anderson1956}. Moderate exciton-phonon interactions may help the exciton to surpass energetic barriers, at the expense of destruction of coherence among different molecular excited states in the aggregate. In the Holstein model, the localized states are linearly coupled to the bath's coordinate (see $\hat{H}_{\rm ex-ph}$). {A rough physical picture would be to interpret exciton-phonon scattering processes as a leak of information of the exciton's state into the environment, destroying coherences among excitonic states and leading to localization of the wave functions in the site basis}. In disordered networks, there exists an optimal value of the exciton-phonon coupling which enhances exciton diffusion and prevents localization. {Coherent localization due to disordered networks will be loosen by scattering processes. Dynamical resonances among sites are created and extended exciton states are formed. If the exciton-phonon interaction is too strong, dynamical localisation of the excitonic polaron takes place}. Quantum transport in disordered networks subject to a dephasing environment has been investigated thoroughly \cite{Madhukar1977, Stern1990,Bonci1996,Sirker2009,DErrico2013a,Moix2013}.

In the following we will study the dynamics of two different variables which will help us to elucidate the distinction between coherent and incoherent transport as a function of the exciton-phonon coupling strength and the degree of disorder in the aggregate. Both measures are functions of the reduced density matrix of the exciton:
\begin{equation}
\hat{\rho}_{\rm nm}=\Tr[\ket{D_1(t)} \bra{D_1(t)} \hat{a}_{\rm n}^{\dag}\hat{a}_{\rm m}]
\label{eq:rho}
\end{equation}
Substituting the $\rm D_{1}$ \textit{ansatz} (Eq.~\ref{eq:D1}) into Eq.(\ref{eq:rho}):
\begin{equation}
\hat{\rho}_{\rm nm}=\alpha^*_{\rm n}\alpha_{\rm m}S_{\rm nm}
\end{equation}
where $S_{\rm nm}$ is the Debye-Weller factor which encompass the dephasing effects of the exciton-phonon interaction and is given by:
\begin{eqnarray}
S_{\rm nm}(t) &=& {\rm\exp}\bigg[ \sum_{\rm q} \big(\lambda^{\ast}_{\rm nq}(t)\lambda_{\rm mq}(t) \nonumber\\
&-&\frac{1}{2}\vert\lambda_{\rm nq}(t)\vert^2-\frac{1}{2}\vert\lambda_{\rm mq}(t)\vert^2 \big) \bigg]
\end{eqnarray}
{\color{black}
\noindent where the variables $\lambda_{\rm nq}~ (\lambda_{\rm mq})$ characterize the coherent state of the bath in the $D_1$ \textit{ansatz} (Eq. \ref{eq:D1}) related to the displacement of the oscillator $\rm q$ at site $\rm n ~(\rm m)$. Clearly $S_{\rm nn}=1$ and the diagonal elements of $\hat{\rho}_{\rm nn}=|\alpha_{\rm n}|^2$ corresponds to the exciton populations in the site basis, while the off-diagonal elements or coherences get generally suppressed by the presence of the Debye-Weller factor $S_{\rm nm}$.}  First, we will analyze the coherence size as a measure of the amount of coherence among different pigments in site {basis}. This measure receives contributions of both the diagonal (populations) and off-diagonal (coherence) elements of the reduced density matrix. The coherence size $L_c(\hat{\rho})$ describes thus the size of a domain within which the chromophores interact coherently. The definition of $L_c(\hat{\rho})$ in terms of the reduced density matrix (\ref{eq:rho}) is \cite{YangZhao1999JPCB}:
\begin{equation}\label{eq:csize}
L_c(\hat{\rho})\equiv \frac{\left(\sum_{\rm nm}|\hat{\rho}_{\rm nm}|\right)^ {2}}{{N}\sum_{\rm nm}|\hat{\rho}_{\rm nm}|^{2}}
\end{equation}
where $N$ is the {number} of total pigments in the ring. In addition to a measure of quantum coherence among chromophores, we want to characterize a measure of localization solely based on the information of exciton population at each site, regardless of coherences. In analogy with the inverse participation ratio (Eq.~\ref{eq:IPR}), we define the \textit{inverse population ratio}:
\begin{equation}\label{eq:IPRrho}
 {\rm IPR}(\hat{\rho})=\frac{1}{\sum_{\rm n} |\hat{\rho}_{\rm nn}|^4}
\end{equation}
Let us consider three extreme cases to emphasize the differences between the IPR$(\hat{\rho})$ as defined in Eq.~\ref{eq:IPRrho} and the coherence size of Eq.~\ref{eq:csize}. Firstly, let us consider a fully localized state $\ket{\psi}=\ket{m}$ in a certain site $m$. The corresponding density matrix is: $\hat{\rho}=\ket{m}\bra{m}$. The IPR$(\hat{\rho})$ in this case is $1$ and the coherence size is $L_c(\hat{\rho})=1/N$. Secondly, we consider a purely statistical mixture without any coherence, then $\hat{\rho}$ is diagonal with elements $1/N$. In this case the IPR$(\hat{\rho})=N$ and the coherence size $L_c(\hat{\rho})=1$. Finally, if we have a maximally coherent state, where $\hat{\rho}_{\rm nm}=1/N ~ \forall ~\rm n,m$ then the IPR$(\hat{\rho})$ is still $N$ as in the second case but the coherence size is now $L_c(\hat{\rho})=N$.

The first two columns of Figure \ref{fig:csizedynamics} show the coherence size dynamics as a function of disorder and exciton-phonon coupling strength. At $t=0$, the coherence size is equal to $1/16$ because the exciton is completely localized in one site. Subsequently the coherence size increases due to the superposition of molecular excited states. {Oscillations appear after the first period of oscillation around $20$ fs. Following a transient signal,} the coherence size reaches an steady state value depending on the exciton-phonon coupling and the strength of disorder. As expected, the steady-state values of the coherence sizes decrease with stronger exciton-phonon coupling and stronger static disorder. During the initial instants ($t<30$ fs), an increase in the Huang-Rhys factor slows down the growth of the coherence size. At later times, the value of the coherence size stays around 2-3 sites below in the former case. Concerning the differences between diagonal and off-diagonal disorder (first and second column), we clearly see that the system with diagonal disorder presents significantly lower values of the coherence size for moderate static disorder ($\sigma_{\rm E}$, $\sigma_{\rm J}$=300, 500, 700 cm$^{-1}$) while remaining very similar in the weakly disordered case of $\sigma_{\rm E}$, $\sigma_{\rm J}$=100 cm$^{-1}$. This can be understood because when $\sigma_{\rm J}$ is close or of the same order as the nearest neighbors coupling ($500$ cm$^{-1}$), various sites will be strongly coupled together and will interact coherently, increasing thus the coherence size.

The last two columns of Figure \ref{fig:csizedynamics} show the dynamics of the inverse population ratio IPR$(\hat{\rho})$ as a function of disorder and exciton-phonon coupling strength. In contrast to the coherence size $L_c(\hat{\rho})$, the IPR$(\hat{\rho})$ is maximum for the case of weak to moderate coupling to the nuclear motion. In the off-diagonal case, the  IPR$(\hat{\rho})$ at later times, seems to benefit of exciton-phonon coupling for all couplings $S>0$. Even quite strong coupling $S = 1.5$ reaches a higher IPR than the uncoupled case $S = 0$. {Many regimes are possible with both coherent and incoherent features, measured by different pairs of $L_c(\hat{\rho})$ and  $IPR(\hat{\rho})$.} In the case of diagonal disorder, the IPR$(\hat{\rho})$ is only enhanced for weak-moderate coupling (S = 0.5-1.0).

Both the coherence size $L_c(\hat{\rho})$ and the inverse population ratio $IPR(\hat{\rho})$ reach an steady state value after $\sim$100 fs. We have investigated the behavior of the average steady state coherence (ASSC) and average steady state IPR (ASS-IPR) versus the disorder strength $\sigma/J$ and $S$ where $J$ denotes the average value of the inter-dimer and intra-dimer coupling which is 542 cm$^{-1}$. The considered values of disorder are therefore $\sigma/J=0.18, 0.55, 0.92, 1.30$ for $\sigma=100, 300, 500, 700$ cm$^{-1}$ respectively. Anderson localization typically occurs for $\sigma/J\sim1$ \cite{Anderson1956} in a nearest-neighbour model. %{When long-range interactions are included localization effects are largely suppressed.}
 A bird's-eye view of both the ASSC and the ASS-IPR are shown in Figures \ref{fig:ASSC}, \ref{fig:ASSIPR} respectively. Diagonal disorder is given by the blue mesh while off-diagonal disorder is given by the red one. The tendency of the ASSC (Fig.~\ref{fig:ASSC}) agrees with our previous discussion: without exception, the ASSC decreases with both increasing values of the exciton-phonon coupling and the disorder strength. It is notorious that the ASSC remains quite high (4-5 sites) even in the case when $\sigma=700$ cm$^{-1}$. In contrast, the ASS-IPR (Fig.~\ref{fig:ASSIPR}) exhibits lower values in the case of zero coupling (S = 0) to the phonon environment due to a heavily localised spectrum. The unusually large value of 12 chromophores for weak disorder and very strong coupling S = 1.5 is not necessarily beneficial for transport as we will see in later sections. In {\text{red}comparison with} a realistic system, where $S = 0.5$ \cite{Damjanovic2002} and $\sigma\sim 100-200$ cm$^{-1}$ \cite{Wu1997,VanOijen1999,Jang2001}, our results show a good compromise between sufficient coherence and exciton delocalization encompassing around 7 sites of the aggregate in agreement with other estimates based on experimental data \cite{Kuhn1997}. {Other dephasing mechanism arising from solvent effects not taking here into account will reduce slightly the amount of delocalization.}

In closed, purely excitonic systems, delocalization is unequivocally related to coherence between the chromophores in the aggregate. When the exciton is embedded in a dephasing environment, the delocalized nature of the exciton picture is perturbed and mixed states with a broad distribution of the population may now emerge in the absence of coherence. The later won't benefit from the interference effects that enhance and suppress different excitation transfer paths which are thought to influence dramatically the transfer efficiency \cite{Olaya-Castro2008}.  Nonetheless, fully coherent dynamics presents strict destructive interference and noise processes that can be characterized by the presence of invariant subspaces of the Hamiltonian leading to exciton trapping \cite{Caruso2009}. As we will see in the next section, this phenomena may also be understood by the existence of subradiant states which remain trapped in the network albeit being dissipative. }

Due to the dephasing nature of the phonon environment, we have seen how the coherence size $L_c(\hat{\rho})$ steadily decreases with increasing exciton-phonon coupling strength. $L_c(\hat{\rho})$ also decreases for an increasing disorder because the quasi-momentum ceases to be a good quantum number and the symmetry of the excitonic spectrum is lost. Differently, the inverse population ratio $IPR(\hat{\rho})$ in disordered aggregates will increase in the presence of weak coupling to the vibrational motion as compared to no exciton-phonon interaction at all, because the noisy environment suppresses the disordered-induced localization phenomena, which relies on wave coherence. By comparing the coherence size and the inverse population ratio we can explore different regimes that interpolate between the coherent and incoherent regimes of transport.

\subsection{Excitation Transfer and Energy Relaxation}
Due to the superradiant nature of the bright states in the LH2, the efficiency of energy transfer is found to be highly dependent on the initial state \cite{Olaya-Castro2008}. As an example, for a tight-binding ring in the presence of uniform and coherent coupling $\gamma$ to a single dissipative channel, only the superradiant state has a non-zero decay width $\Gamma=N\gamma$, being $N$ the total number of chromophores in the ring \cite{Celardo2014}. If the exciton is required to be transferred along this channel, the presence of superradiance enhances dramatically the transfer rate as compared to the decay width of an isolated site, a phenomena called supertransfer \cite{Strek1977,Abasto2012,Strumpfer2012,Kassal2013}, a corollary companion of superradiance. Superradiance in cyclic aggregates and its robustness to static disorder has been extensively studied \cite{Meier1997,Monshouwer1997,Chernyak1999,YangZhao1999JPCB,Celardo2014}. The even number of pigments in the LH2 ring provides constructive interference between the left-travelling and right-travelling wave, which has been shown to enhance the excitation transfer in LH2 assemblies \cite{Ye2012}. In this work we will investigate excitation transfer to a dissipative channel for different initial conditions and regimes of static disorder and exciton-phonon coupling strength.
\begin{figure}[htb!]\label{fig:cEnhnc}
  \includegraphics[scale=0.55]{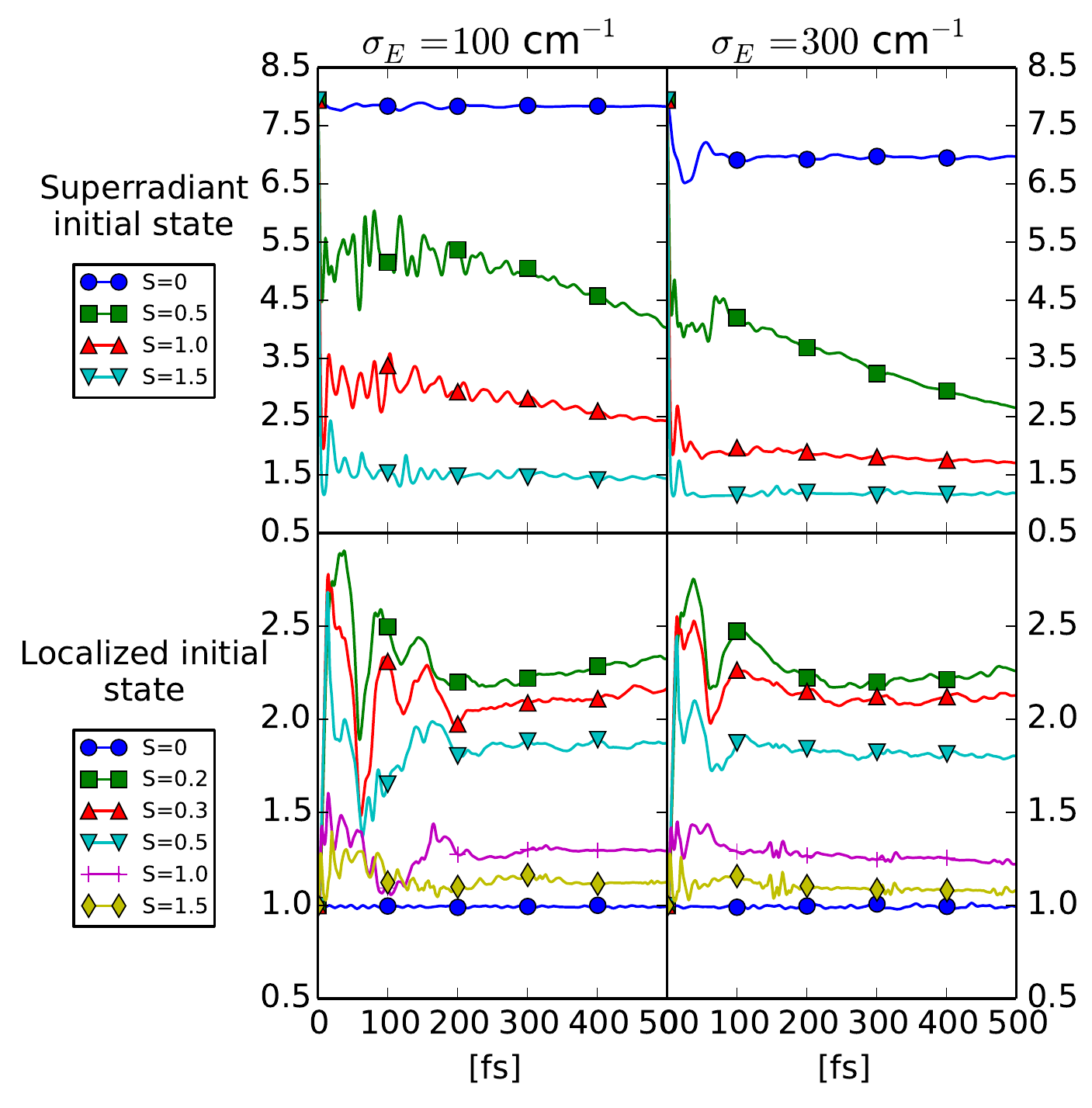}\\
  \caption{Superradiance enhancement factor $L_{s}(\rho)$ (in units of the radiative decay rate of a single chromophore $\gamma$) as a function of exciton-phonon coupling and static disorder (left and right column, diagonal disorder with $\sigma_{\rm E}=100,300$ cm$^{-1}$ respectively). Two initial conditions are considered: a superradiant state $\rm k=-1$ with a radiative decay (and oscillator strength) of 8 pigments and an initially localized excitation at the center of the ring.}
\end{figure}
\begin{figure}[htb!]\label{fig:Psink}
  \includegraphics[scale=0.55]{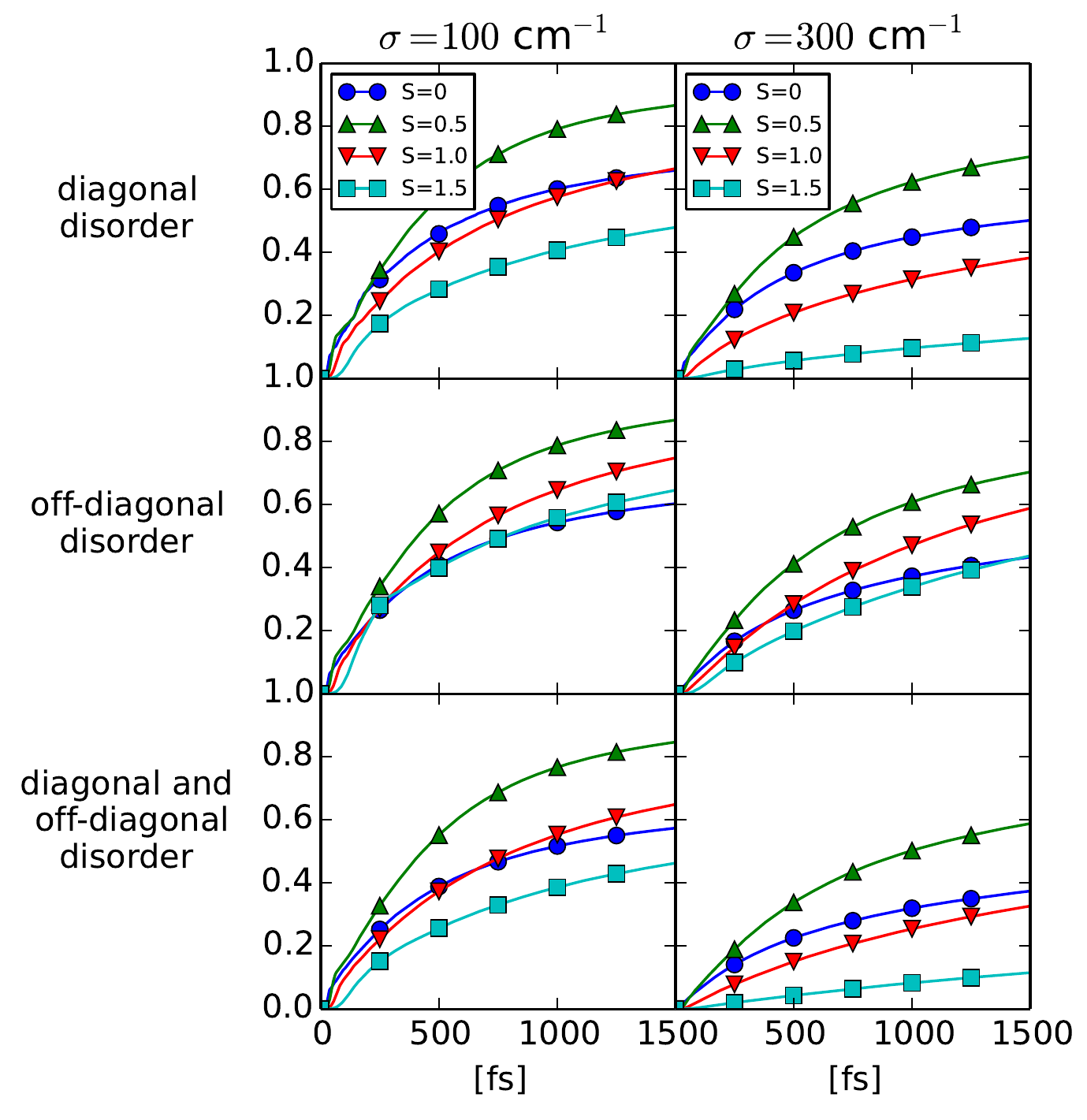}\\
  \caption{Probability at the sink ${\rm P}_{\rm sink}(t)$ as a function of the exciton-phonon coupling $S$ for an initially localized excitation (site 8, bath at the ground state) exiting at the opposite site of the ring with fixed diagonal disorder (upper panel $\sigma_{\rm E}$), off-diagonal disorder (middle panel,  $\sigma_{\rm J}$) and both diagonal and off-diagonal disorder (lower panel $\sigma_{\rm E}=\sigma_{\rm J}$). The dissipation rate is set to $\gamma=0.1~\omega_0$ and $Q_{\rm nm}=\delta_{\rm n0}$. }
\end{figure}
All information about excitonic superradiance (along with the system's geometry) \cite{Gross1982} is contained within the reduced density matrix. The efficiency of excitation transfer is intimately related to the superradiance enhancement factor, which is defined in terms of the reduced density matrix and the orientation of dipoles in the system:

\begin{equation}\label{eq:cohEnhnc}
L_s(\hat{\rho})=\sum_{\rm n,m} \boldsymbol{\mu}_{\rm n} \cdot \boldsymbol{\mu}_{\rm m}~ \rho_{\rm nm}=\frac{\Gamma(\hat{\rho})}{\gamma}
\end{equation}
\noindent where $\gamma$ is the radiative decay of a single pigment and $\boldsymbol{\mu}_{\rm n}$ is the transition dipole moment of chromophore $n$. The superradiance enhancement factor $L_s(\rho)$ measures the radiative decay rate of an aggregate, given by $\Gamma(\rho)$ in Eq. (\ref{eq:cohEnhnc}), in units of the radiative decay rate of a single chromophore $\gamma$. We remind that the clean, homogeneous system presents two superradiant, degenerate states neighbouring the bottom of the exciton band. Each of them carries an oscillator strength of $N/2$, in our case 8 dipoles because there are 16 pigments in our system. These superradiant states {each presents} a superradiance enhancement factor of 8 dipoles: their decay rate is 8 times faster than that of a single chromophore. We have plotted in Figure \ref{fig:cEnhnc} the dynamics of $L_s(\hat{\rho})$ for different initial conditions. If the initial state corresponds to one of the superradiant states neighbouring the bottom of the exciton band, exciton-phonon interactions will slow down the decay rate and reduce the superradiance coherence factor. It may be argued that the bright states of the disordered Hamiltonian do not correspond to those of the homogeneous one but as we can see in Figure \ref{fig:cEnhnc} for the case S = 0, the superradiance enhancement factor is still very close to $N/2=8$ dipoles, and we may assume that this state is robust to the presence of realistic values of static disorder. For an initially localized state we see that the environmental phonons enhance the radiative decay. We expect therefore that when the disordered ring is open to a population sink, an initially localized excitation will exit the network faster in the presence of some environmental dephasing, due to the decay rate enhancement.
\begin{figure}[htb!]\label{fig:nk}
  \includegraphics[scale=0.38]{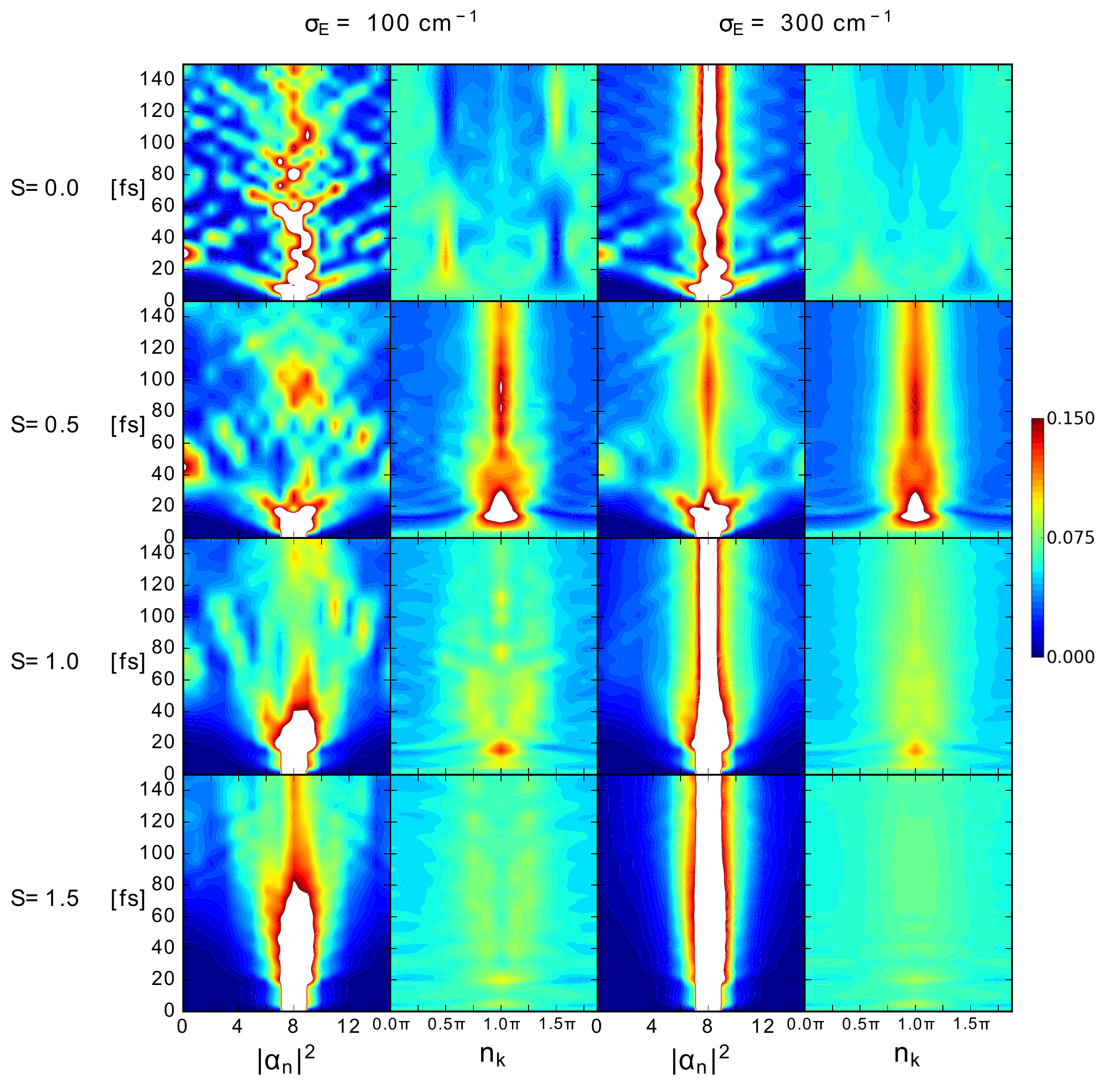}
  \caption{Exciton transfer in site space (1st and third columns) and momentum space (2nd and 4th columns) for various exciton-phonon coupling strengths (from top to bottom: S = 0, 0.5, 1.0, 1.5) and diagonal disorder strengths ($\sigma_{E}=100,300$ cm$^{-1}$ for left, right panels respectively). The dissipation rate is set to $\gamma=0.1~\omega_0$ and $Q_{\rm nm}=\delta_{\rm n0}$. }
\end{figure}
\begin{figure}[htb!]
  \centering
  \includegraphics[scale=0.45]{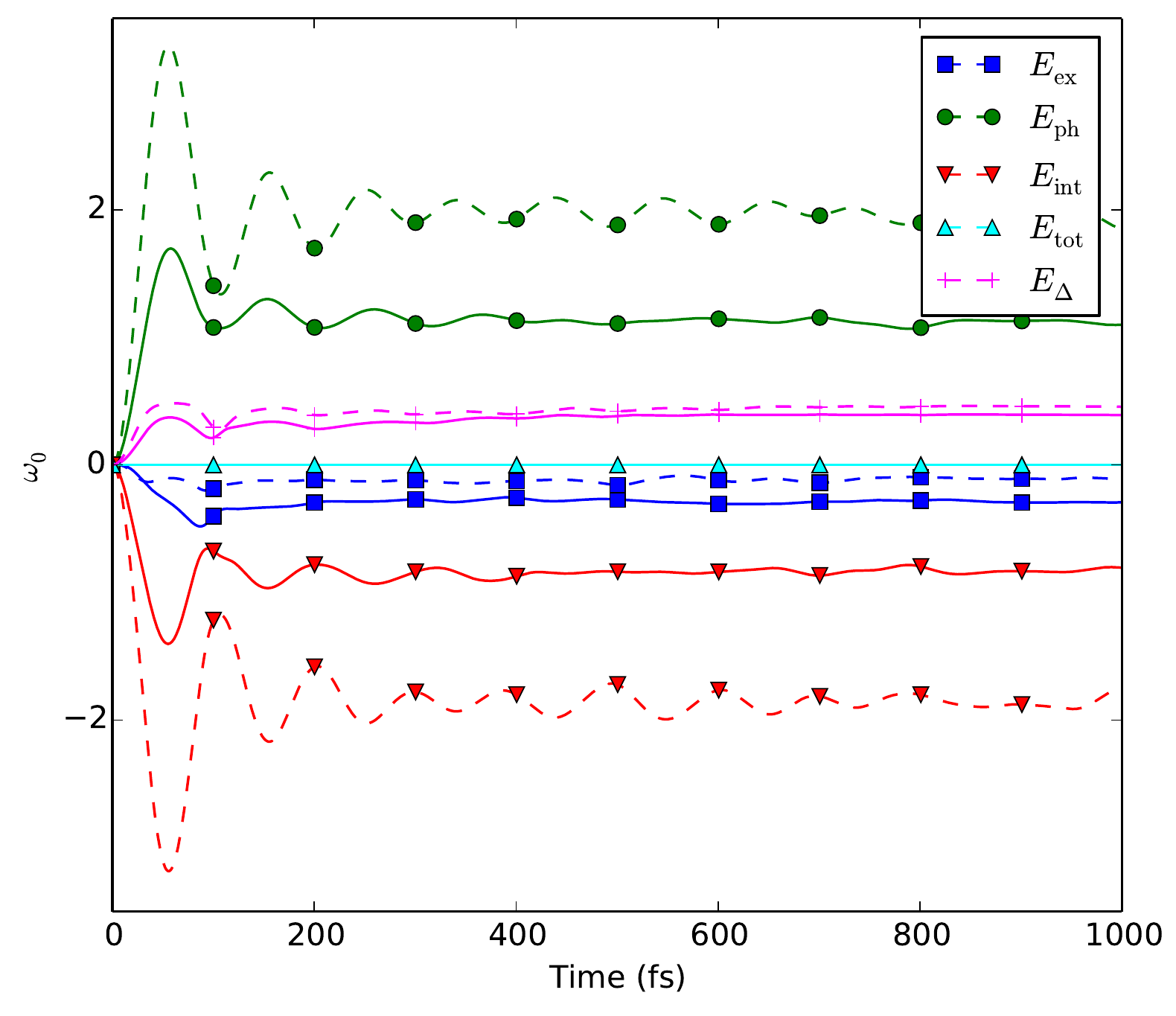}
  \caption{Energy dynamics in the B850 band with $\sigma_{\rm E}=$100 cm$^{-1}$ for an initially localized state. Simulation parameters are a phonon bandwidth of $W=0.5$ and an exciton-phonon coupling of $S = 0.5$ (solid curves) and $S = 1$ (dashed curves). The expressions for the exciton energy (squares), bath's energy (circles) and the interaction energy (down-triangles) can be found in Appendix B (Eqs.~\ref{eq:E_ex}, \ref{eq:E_bath}, \ref{eq:E_int}). A quantity denoted ``deviation amplitude'' $\Delta(t)$	 (Eq. \ref{eq:devamplitude}, plus sign markers) is introduced as a variable which measures the optimality of the variational \textit{ansatz}  (see Appendix).  The total energy is indeed conserved (up-triangles). All these quantities are in units of the phonon characteristic frequency $\omega_0$=1670 cm$^{-1}$. The on-site energy  $\epsilon_{{n}}$ has been subtracted from $E_{\rm{ex}}$ and $E_{\rm{tot}}$.}\label{fig:energydynamics}
\end{figure}

We will model exciton transfer by attaching our ring to a single dissipative channel, by means of an effective non-hermitian Hamiltonian which will leak probability out of the single-exciton manifold. We thus substitute the excitonic term (Eq. \ref{eq:Hex}) of the complete Hamiltonian (\ref{eq:Hamiltonian}) by:
\begin{equation}\label{eq:Hexeffective}
\hat{H}_{\rm{eff}}=\hat{H}_{\rm{ex}} - i\frac{\gamma}{2} \hat{Q}
\end{equation}
\noindent where the operator $\hat{Q}$ determines the structure of the coupling between the electronic system and the dissipative channel \cite{Celardo2014}. This new dissipative term in the Hamiltonian  induces a non-conserving term in the evolution of the excitonic amplitudes:
\begin{equation}\label{eq:alphaeff}
\dot{\alpha}_{\rm n}(t)=\frac{i}{\hbar}\left( T_{\rm n} + \alpha_{\rm n} R_{\rm n} + \frac{i\gamma}{2} \sum_{\rm m} \alpha_{\rm m} Q_{\rm nm} S_{\rm nm}\right)
\end{equation}
\noindent where the operators $T_{\rm n}$ and $R_{\rm n}$ are the norm-conserving terms defined in the appendix and $S_{\rm nm}$ is the Debye-Weller factor. Dynamic and static disorder will disrupt the invariant subspaces of the Hamiltonian that can lead to exciton trapping \cite{Caruso2009}.
%The case for $Q$ diagonal, $Q_{\rm nm}\propto \mathbb{1}_{\rm ex}$ (identity in the single-exciton manifold) corresponds to local-dissipation, whereas the case with $Q_{\rm nm}=1\quad\forall \,\rm n,m$ receives the name of ``coherent dissipation'' \cite{Celardo2014}).

Let us consider a case where a local excitation is transferred to the opposite side of the ring. We create an excitation at site $8$ and connect the opposite site to a population sink $Q_{\rm nm}=\delta_{\rm n0}$. We have plotted in Figure \ref{fig:Psink} the probability of successful transfer to the target sink for various values of disorder. In concordance with the previous analysis of the superradiance enhancement factor, we observe that the excitation transfer efficiency is maximized by the presence of weak to moderate environmental phonons as compared to the disordered bare exciton case. In particular, when $S = 0.5$ (green triangles) the transfer efficiency is in any case superior to that of the uncoupled, disordered ring. We can understand this environmentally assisted transport in terms of exciton dynamics in the momentum basis of the original, non-disordered exciton Hamiltonian. The population of electronic momentum $k$ under the influence of the dephasing environment can be computed in the following way by means of the reduced density matrix (Eq.~\ref{eq:rho}):
 \begin{equation}
n_{\rm k}=\langle \hat{a}_{\rm k}^\dag \hat{a}_{\rm k}\rangle= \frac{1}{N}\sum_{\rm n,m} e^{i\rm k(n-m)} {\rho}_{\rm nm}
\end{equation}
%\begin{eqnarray*}
%n_k&=&\langle a_k^\dag a_k\rangle=	\frac{1}{N}\sum_{n,m} e^{ik(n-m)} \langle a_n^\dag a_m \rangle \\\nonumber
%& =& \frac{1}{N}\sum_{n,m} e^{ik(n-m)} \Tr\left[\rho a_n^\dag a_m \right]   \\\nonumber
%& =& \frac{1}{N}\sum_{n,m} e^{ik(n-m)} \Tr\left[\ket{D_1}\bra{D_1} a_n^\dag a_m \right]\\\nonumber
%& =& \frac{1}{N}\sum_{n,m} e^{ik(n-m)} \alpha_j^* \alpha_l S_{nm}   \\\nonumber
%& =& \frac{1}{N}\sum_{n,m} e^{ik(n-m)} \tilde{\rho}_{nm}= \left( M_{\rm DFT} ~\tilde{\rho} ~ M_{\rm DFT}^{-1}\right)_{kk}   \\\nonumber
%\end{eqnarray*}
Figure \ref{fig:nk} shows the exciton populations in the site and momentum basis when the ring is opened with the same dissipation parameters of Figure \ref{fig:Psink}. Rows correspond to different exciton-phonon coupling strenghts (S = 0, 0.5, 1.0, 1.5). The first two columns refer to diagonal disorder with $\sigma_{E}=100$ cm$^{-1}$ and the rest to $\sigma_{E}=300$ cm$^{-1}$. Phase noise in the site basis becomes amplitude noise in the excitonic { basis} and a relaxation mechanism is enabled under the action of the phonon environment \cite{HuelgaVib}. We clearly observe that when the exciton interacts with the phonon environment, the exciton rapidly relaxes at the bottom of the exciton band ($k=\pi$) \cite{Note1}, which corresponds to those states with the highest delocalization properties and belong to the chaotic, diffusive regime, as discussed in Section III-A. This result is consistent with recent pump-probe experiments on the LH2 where it was reported that exciton relaxation is completed in 300 fs with a steady-state wave packet where the exciton resides in the three lowest energy states \cite{Novoderezhkin2011b}. The distribution of population in different momentum states gets broader with an increased exciton-phonon coupling strength, leading to exciton localization. Li \textit{et al.}~\cite{Li2015} ascribe this enhancement of efficiency to a ``rejuvenation'' of momentum states with a high group velocity. Similar results are obtained for off-diagonal disorder (not shown).

Let us look into the dynamics of the energy terms of the Hamiltonian (\ref{eq:Hamiltonian}) in order to emphasize the relaxation mechanism in the exciton manifold (Figure \ref{fig:energydynamics}) when there is no excitonic leakage ($Q_{\rm nm}=0~\forall\rm~ n,m$). The expressions for the exciton energy (squares), bath's energy (circles) and the interaction energy (down-triangles) can be found in the Appendix (Eqs.~\ref{eq:E_ex}, \ref{eq:E_bath}, \ref{eq:E_int}). Furthermore, a quantity denoted ``deviation amplitude'' $\Delta(t)$(eq. \ref{eq:devamplitude}, plus sign markers in Fig.~\ref{fig:energydynamics}) is introduced as a variable which measures the optimality of the variational \textit{ansatz}  (see Appendix). All these quantities are in units of the phonon characteristic frequency $\omega_0$=1670 cm$^{-1}$. We have started with an initially localized excitation and therefore the electronic kinetic energy of the exciton \cite{Dorfner2015} is equal to the excitation energy of a single site, in our case, it corresponds to the excitation energy of the $Q_y$ transition of a BChl a molecule. We have subtracted this energy from $E_{\rm ex}$ and $E_{\rm total}$ for illustration. The total energy is indeed conserved (up-triangles) as guaranteed by our variational formalism. The phonon bath experiences a transient displacement upon exciton injection which induces relaxation of the kinetic energy in the exciton manifold. As we have foreseen in the momentum analysis, this relaxation is more effective in the case of weak exciton-phonon coupling (compare solid blue squares with S = 0.5 and dashed blue squares with S = 1 in Figure \ref{fig:energydynamics}). After a period of non-equilibrium  the exciton energy reaches a steady state value around 250 fs. Both the interaction and the phonon energy oscillate with a frequency dependent on the phonon characteristic frequency and its bandwidth $W$. These oscillations are more persistent in the case of strong exciton-phonon coupling. A detailed analysis of energy relaxation in the Holstein one-dimensional model can be found in Ref.~\cite{Dorfner2015}.

\subsection{Spectroscopy}
% Energy Transfer in LH2 of Rhodospirillum Molischianum, Studied by Subpicosecond Spectroscopy and Configuration Interaction Exciton Calculations

%In this section we will first derive the absorption spectra of the LH2 computed from a time-dependent perspective employing linear response theory \cite{MukamelBook}. A secondary, low-frequency thermal bath that accounts for solvent effects and protein motion is included by means of an overdamped Brownian oscillator model \cite{Gu1993,Li1994,Knox2002,Novoderezhkin2011b}. This thermal bath is specified by three parameters: the reorganization energy $\Lambda_0$, the solvent's relaxation time $\gamma_0^{-1}$ and temperature $T$. In any case this thermal bath will produce an exponential decay of the response function.  Details of the theoretical derivation of the absorption spectra can be found in the Appendix. Figure $\ref{fig:absorption}$ shows the absorption spectra for in the B850 band in the LH2 complex from \textit{Rs. molischianum} with $\gamma_0=200$ cm$^{-1}$, $\lambda_0=200$ cm$^{-1}$ (Eq.~\ref{eq:Drude}) at room temperature ($\rm T=300 K$) with static disorder $\sigma_{\rm E}=\sigma_{\rm J}=100$ cm$^{-1}$ and S=0.7.

%\begin{figure*}[htb!]
%  \centering
%  % Requires \usepackage{graphicx}
%  \includegraphics[width=0.7\linewidth]{ZPL1670.eps}\\
%  \caption{ Absorption spectra with $\gamma_0=35$ cm$^{-1}$, $\lambda_0=100$ cm$^{-1}$ at T=77 K for different diagonal (top panels) and off-diagonal (bottom panels) disorder strengths.}\label{fig:absorption}
%\end{figure*}

\begin{figure}
\includegraphics[width=1.2\linewidth,trim=150 90 0 40]{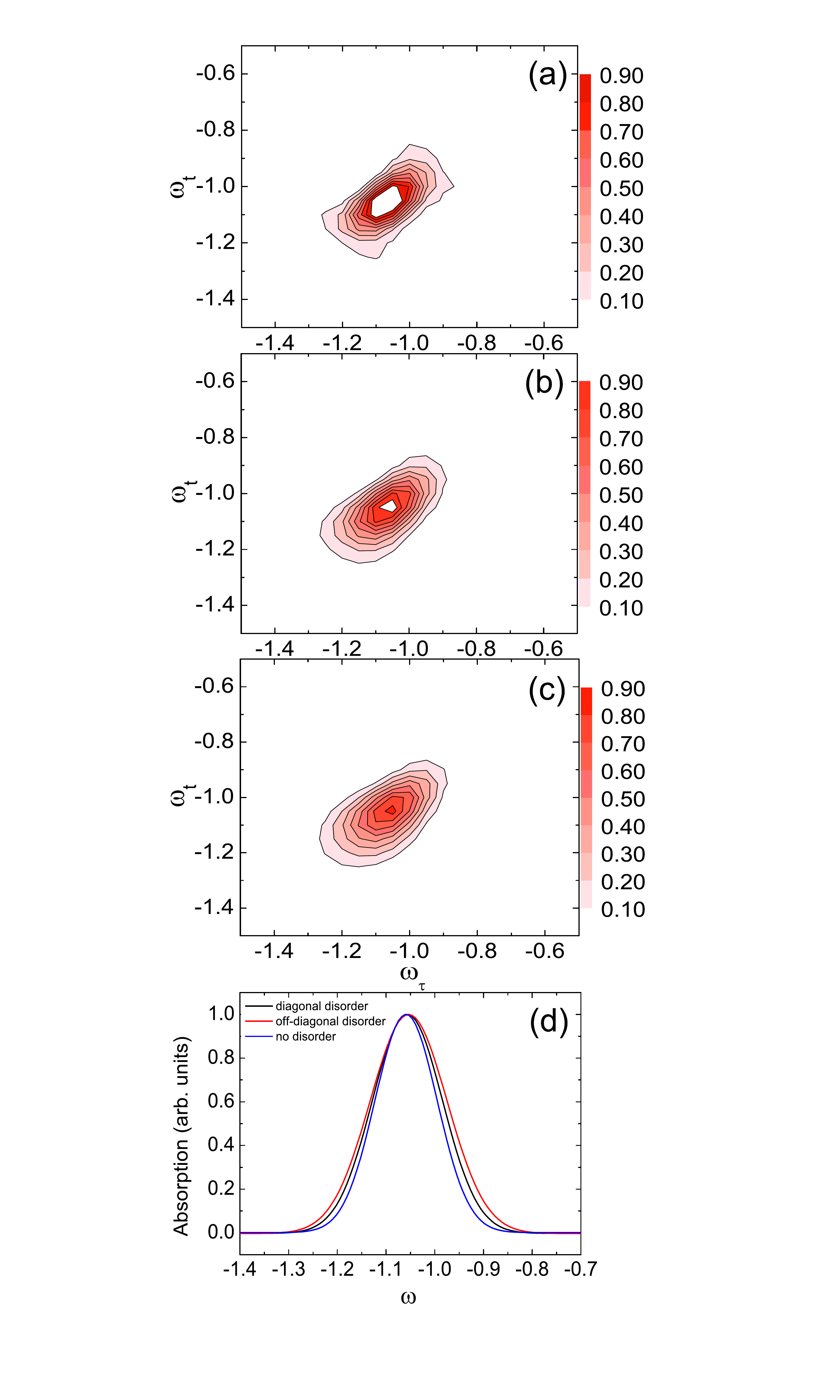}\\
  \caption{2D spectra (sum of the real part of the rephasing and non-rephasing pathways, in units of the phonon characteristic frequency $\omega_0$) with $\gamma_0=35$ cm$^{-1}$, $\lambda_0=100$ cm$^{-1}$, $S=0.5$ and $W=0.5$ at T=77 K for (a) no disorder, (b) diagonal disorder strength $\sigma_E=100$ cm$^{-1}$ and (c) off-diagonal disorder strength $\sigma_J=100$ cm$^{-1}$. The results of (b) and (c) have been averaged over 200 samples for a single ring system. And the waiting time is $T_w=0$. (d)Linear absorption spectra.}\label{2D}
\end{figure}

In this section we will investigate the two-dimensional (2D) photon-echo spectra of the LH2 (Fig.~\ref{2D}) computed from
a time-dependent perspective \cite{MukamelBook}). The details of the derivation are shown in Appendix $\rm B$. Due to
the high efficiency of the $\rm D_1$ Ansatz method, we can easily perform the calculation of 2D spectra even in the
presence of diagonal (off-diagonal) disorder \cite{Huynh,2DD1}.  A secondary, low-frequency thermal bath that accounts
for solvent effects and protein motion is included by means of an overdamped brownian oscillator
model \cite{Gu1993,Li1994,Knox2002,Novoderezhkin2011b}. This thermal bath is specified by three parameters:
the reorganization energy $\Lambda_0$, the solvent's relaxation time $\gamma_0^{-1}$ and temperature $T$.
The case without disorder is plotted in Fig.~\ref{2D}(a). The main peak corresponding to the zero phonon line
transition is located at $\omega_{\tau}=\omega_t\approx-1.08~\omega_0$ (in units of the phonon characteristic
frequency $\omega_0$). The monomer transition energy has been set to zero and therefore the peak's shift from
zero is due to the reorganization energy and excitonic coupling between monomers. The 2D spectra of the LH2 is
very similar to that of a two-level system, since a doubly degenerate bright state dominates the absorption at
the bottom of the B850 band and the weak exciton-phonon coupling $S=0.5$ is not strong enough to show a clear
vibronic progression. When the system is not disordered, the peak shape is elliptical and homogeneously broadened.
Both diagonal and off-diagonal disorder (Figs.~\ref{2D}~b, c, respectively) round the elliptical features of
the non-disordered spectra and elongate them along its diagonal.
Both diagonal and off-diagonal disorder give rise to similar 2D spectra, although the later shows slightly stronger broadening for the values
considered ($\sigma_E=\sigma_J=$100 cm$^{-1}$). The linear absorption spectra can be read out from the diagonal
($\omega_{\tau}=\omega_t$) part of the 2D spectrum at $T_w=0$, as shown in Fig.~\ref{2D}(d). The off-diagonal disorder case
indeed exhibits a stronger inhomogeneous broadening in comparison to the other two cases.
The experimental 2D spectra of LH2 rings shows energy transfer between
the B850 and B800 bands  \cite{EngelLetters,EngelPNAS2012}, being the former correctly {characterized} by our model.
Low-lying energy dark states in the LH2 have been {hypothesized} to constitute charge transfer states and polaron pairs that
may play an important role in the dynamics of energy transfer \cite{LH2Dark}.
%The dephasing effect should be described by the phase uncertainty in the wave amplitude $\alpha_n$, which could be caused not only by the evolution of the wave function in the pigment network, but also by the environment~\cite{Xiong}.
%Actually, the static disorder can also produce additional uncertainties to the system.
%The introduction of diagonal static disorder (Fig.~{2D}(b)) elongates the main feature in the anti-diagonal direction.
%First, we observe that the peaks become inhomogeneously broadened for the diagonal(or off-diagonal) disorder case by comparing Fig.~\ref{2D}(a) and (b)(or (c)). Secondly, from Figs.~\ref{2D}(b) and (c), it is found that the disorder also induce a homogeneous broadening in the low frequency region, which suggests the disorder indeed causes the uncertainties of the wave amplitude and equivalently shows the dephasing effect.

\subsection{Exciton Transport}

Quantum transport in noisy networks has received much attention \cite{Madhukar1977, Stern1990,Bonci1996,Sirker2009,DErrico2013a,Moix2013}. In the case of a free particle, wave coherence leads to ballistic transport with the mean squared displacement scales with time squared $\langle x^2(t)\rangle\sim t^2$. In the presence of scattering sources the diffusion becomes classical $\langle x^2(t)\rangle\sim t$ at later times. These scattering events may originate from impurities, a disordered energy landscape or environmental phonons. Thus, the environment obtains information about the particle's position by means of scattering processes, a phenomenon denominated noise-induced diffusion \cite{DErrico2013a}. When the initial state corresponds to a fully localized excitation, we can measure the mean squared displacement (MSD) from the origin:
\begin{eqnarray}
\label{eq:MSD}
{\rm MSD}(t)&=& \sum_{\rm n}   \rho_{\rm nn}(t)  (x(t)-x_0)^2\\\nonumber
&=& \sum_{\rm n} (d_{\rm n} )^2 \rho_{\rm nn}(t) \sim  D t^{\gamma} \nonumber
%{\rm MSD}(t)&\sim & D t^{\gamma}
\end{eqnarray}
\noindent where $d_{\rm n}$ is the distance between the initial site and site $\rm n$ and the parameters $D$, $\gamma$ are the mobility coefficient and diffusion exponent parameters, respectively, that characterize the diffusive behavior of the exciton transport during the initial evolution.

\begin{figure}[tbh!]
  \centering
  \includegraphics[width=1.02	\linewidth]{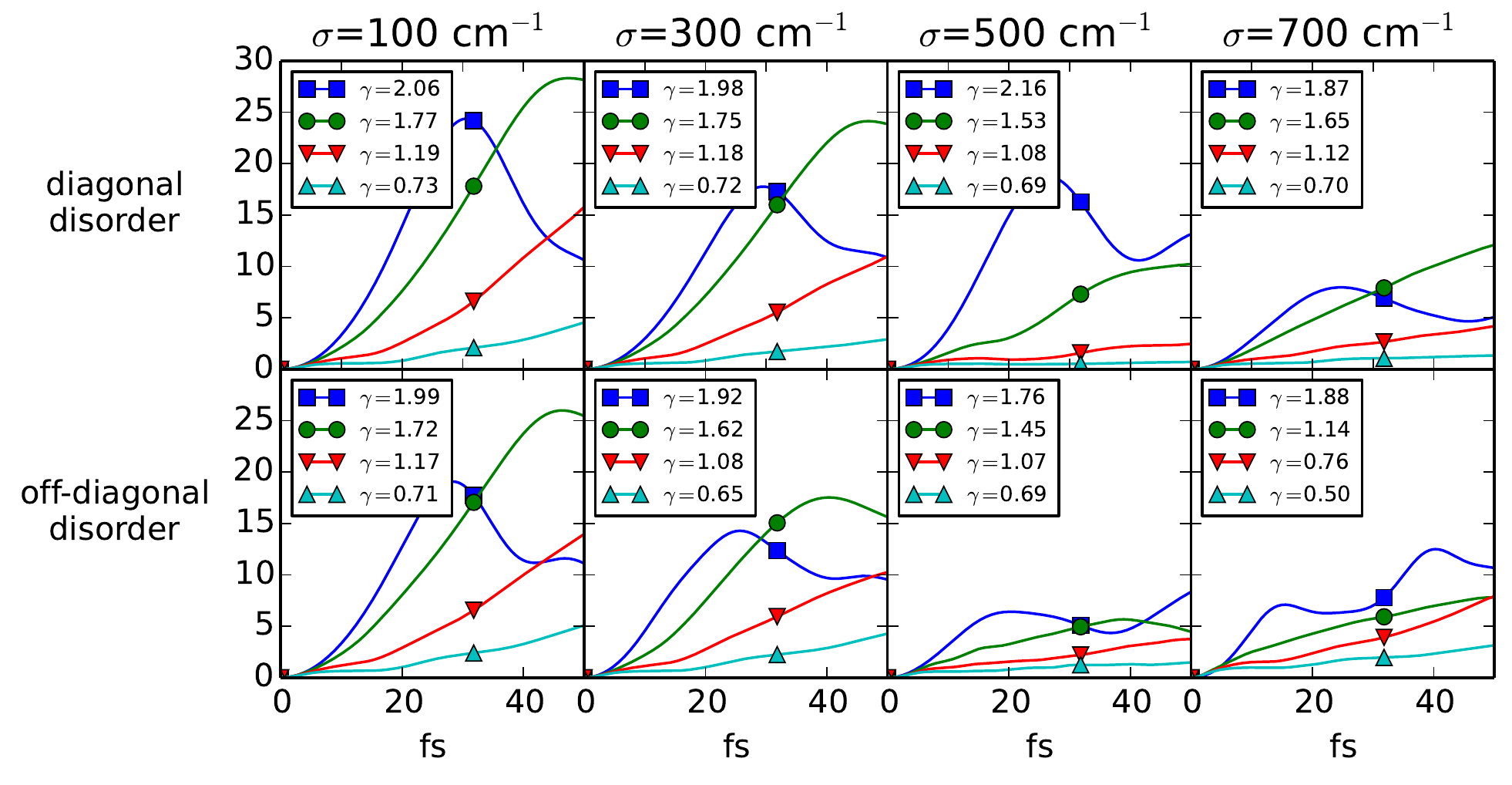}
  \caption{Mean squared displacement (MSD, in units of nm$^{2}$) during the first 50 fs for diagonal disorder (first row) and off-diagonal disorder (second row) as a function of exciton-phonon coupling (blue squares S = 0, green circles S = 0.5, red down-triangles S = 1.0, cyan up-triangles S = 1.5) and disorder strength (columns). The MSD diffusion exponent $\gamma$ as defined in Eq.~\ref{eq:MSD} is shown correspondingly at each legend where the fit has been taken at the interval 1.6 fs $<$ t $<$16 fs.}\label{fig:msdt}
\end{figure}

We plot in Figure \ref{fig:msdt} the mean squared displacement during the first 50 fs for various forms of static disorder and exciton-phonon coupling (blue squares S = 0, green circles S = 0.5, red down-triangles S = 1.0, cyan up-triangles S = 1.5). We have extracted the diffusion exponent $\gamma$ during the first 16 fs.  These curves present an initial growth of the MSD with a slope that is dependent on the exciton-phonon coupling strength. When the exciton explores the totality of the ring, a maximum value is reach. The time required for the exciton to reach its maximum MSD is dependent on the Huang-Rhys factor S. For example, for $\sigma_{\rm E}=100$ cm$^{-1}$, the maximum is reach at 25 fs for S = 0, 40 fs for S = 0.5, 60 fs for S = 1.0 and 120 fs for S = 1.5. If the exciton-phonon value is strong and the static disorder too, the MSD remains with a low value as an indication of exciton localization or trapping. As expected, the diffusion exponent is $\gamma\sim 2$ when the exciton-phonon interaction is shut down (blue squares, S = 0), and it is significantly decreased in the presence of the dephasing environment. In particular, when the interaction with the phonon environment is too strong (cyan-up triangles, S = 1.5), transport becomes sub-diffusive ($\gamma < 1$) as a result of dynamic localization.

\section{Conclusion}

In this work we have investigated quantum transport in disordered nanorings embedded in a phonon environment. The B850 band of the LH2 complex from \textit{Rs. molischianum} with long-range interactions and anti-parallel dipoles has been employed as a prototype \cite{Koepke1996}.  First, we have studied the localization properties of a disordered nanoring by analyzing the spectral statistics of their Hamiltonians finding that localization is highly dependent on energy. {In particular, thanks to the specific geometry of the LH2, extended states are found at low energies even at high values of disorder. These states, in assistance of intra-molecular vibrations, dominate the relaxation dynamics in the LH2.} We monitor the dynamics of these vibronic excitons as a function of various forms of static disorder and exciton-phonon coupling strength, finding {an optimal, exciton-phonon coupling ($S\sim0.5$, 835 cm$^{-1}$) that facilitates exciton transfer in the presence of disorder}. We corroborate these findings by analyzing several measures of delocalization which ease the separation of coherent and incoherent transport regimes. According to our results, { the average coherence length within our realistic model of the LH2 extends over 7 pigments, neglecting additional thermal dephasing that will further decrease this value by a few sites \cite{Meier1997,YangZhao1999JPCB}.} In addition, we have studied the case of a disordered ring coupled to a dissipative channel and verified that indeed excitation transfer in disordered rings can be enhanced by the presence of environmental phonons, opening up a relaxation mechanism that funnels the exciton to the bottom of the excitonic band. We have utilized time-dependent variational dynamics in combination with the Davydov $\rm D_1$ trial state \textit{ansatz}, and exploited the parallel architecture of Graphic Processor Units (GPU) to perform { our numerical simulations}.

%According to our results,  around 7 pigments in the LH2 form a coherent patch during the excitation transport process, although the addition of thermal effects may further decrease this value \cite{Meier1997,YangZhao1999JPCB}.
%We have addressed in this paper the impact of static disorder on the quantum transport.In this paper, we have investigated the effects of the diagonal disorder and the off-diagonal
%disorder on the excitonic localization by using $\rm D_1$ {\it ansatz} method. The ASSC as a measurement of the coherence is defined in order to study the effects of the various disorder on the excitonic localization. It is found that quantum coherence survives from the environment induced diagonal disorder for a realistic LH2 system, and may robustly maintain a high efficient energy transfer. Usually, a bigger disorder or a stronger exciton-phonon coupling can give rise to the excitonic localization, while the off-diagonal disorder shows a complicated behaviors on it.
\section*{Acknowledgments}
Support from the Singapore National Research Foundation through the Competitive Research Programme (CRP) under Project No.~NRF-CRP5-2009-04 is gratefully acknowledged. This work also was supported by Spanish MINECO project FIS2012-34479 and CSIC project I-Link0938. One of us (K.W.S.) also thanks NNSF of China (Nos.~11404084, ~11574052) for partial support.

\appendix

{
\section{Spectral Statistics}
The repulsion parameter is extracted from the behaviour of the nearest neighbour spacing distribution near degeneracy: $P(s)\sim s^\beta$ $(s \rightarrow 0)$, where $s_i = (E_{i+1} -E_i)$ and $E_i$ are the energies of the Hamiltonian ordered from lowest to highest value.  Integrable systems present a Poissonian profile of the spacing {distribution} with $\beta=0$ while chaotic systems show level repulsion, avoiding degeneracies with $\beta \sim 1$. In order to study the repulsion parameter $\beta$ of spectral statistics, we fit the nearest neighbor spacing distribution to the phenomenological Brody distribution \cite{Brody73} which has been used succesfully for similar purposes in many areas related to localization in disordered systems or quantum chaos \cite{Wilkinson91,Molina01,Flores13,Frisch14,Mur15}:
\begin{equation}\label{eq:repulsion}
 P(s)= A(\beta+1)s^{\beta} \exp({-As^{\beta+1}}). \,\,A=\Gamma\left[\frac{\beta+2}{\beta+1}\right]^{\beta+1}
\end{equation}
where the constant $A$ is needed for proper normalization {and $\Gamma$ is the Gamma function}. For localized systems,
level repulsion is diminished as neighboring levels tend to localize in non-overlapping spatial regions, $\beta=0$ and the nearest neighbor spacing distribution
is equal to the Poisson distribution $P(s)=\exp{(-s)}$.
In systems of disordered but extended states the form of $P(s)$ can be obtained from RMT and can be approximated by the Wigner-Dyson distribution which is equal to the Brody distribution with $\beta=1$, $P(s)=(\pi/2)\exp{(-\pi s^2/4)}$ \cite{Mehta_Book}. The Brody distribution interpolates smoothly between these two limits. {Previous} to the computation of the $P(s)$, the unfolding of the spectra to unit spacing is needed. The unfolding procedure filters the smooth part of the spectrum by performing a local average on the nearest neighbor distances \cite{Gomez02}. Due to the small number of individual levels we need to resort to an ensemble unfolding procedure  \cite{Haake_book}, the average spacing at each energy is taken from the ensemble average. This procedure has been proven to provide correct results for the Anderson disordered model.
}

\section{Derivation of polaron dynamics}

The Lagrangian in the Dirac-Frenkel time-dependent variational approach takes the following form:
\begin{equation}
\mathcal{L}=\bra{D_1(t)} \frac{i\hbar}{2} \left( \overset{\rightarrow}{\frac{\partial}{\partial t}}- \overset{\leftarrow}{\frac{\partial}{\partial t}}\right)-H \ket{D_1(t)}
\label{eq: Lagrangian}
\end{equation}
Terms of the Lagrangian involving time derivatives can be expressed as:
\begin{align}
\bra{D_1}\overset{\rightarrow}{\partial_t}\ket{D_1} = &\sum_{\rm {n}}  \bigg( \alpha_{\rm {n}}^* \dot{\alpha_{\rm {n}}}  \\\nonumber
& +|\alpha_{\rm {n}}|^2 \sum_{\rm {q}} \left(\frac{\dot{\lambda_{\rm {n}{q}}}\lambda_{\rm {n}{q}}^*-\dot{\lambda_{\rm {n}{q}}}^*\lambda_{\rm {n}{q}})}{2} \right) \bigg)\\
\bra{D_1}\overset{\leftarrow}{\partial_t}\ket{D_1} &=\left( \bra{D_1}\overset{\rightarrow}{\partial_t}\ket{D_1} \right)^*
\end{align}
The other term $\bra{D_1(t)} H \ket{D_1(t)}$ corresponds to the expected value of the total energy, which is a conserved quantity and can be split in three contributions:
\begin{align}
\label{eq:E_ex}
\braket{D_1|H_{\rm{ex}}|D_1}& = \sum_{\rm {n}{n}^\prime} \alpha_{\rm {n}}^* \alpha_{\rm {n}^\prime} J_{\rm {n}{n}^\prime}  S_{\rm {n}{n}^\prime}\\
\label{eq:E_bath}
\bra{D_1}H_{\rm{bath}}\ket{D_1}
&= \sum_{\rm {n}}\sum_{\rm {q}} \hbar \omega_{\rm {q}} |\alpha_{\rm {n}}|^2  |\lambda_{\rm {n}{q}}|^2\\\nonumber
\label{eq:E_int}
\bra{D_1}H_{\rm{ex-ph}}\ket{D_1}&=\\
=\sum_{\rm {n}}|\alpha_{\rm {n}}|^2 &\bigg( -\frac{2}{\sqrt{N}} \sum_{\rm {q}} \hbar \omega_{\rm {q}} g_{\rm {n}{q}} \operatorname{Re}(e^{-i	\rm {q}{n}} \lambda_{\rm {n}{q}}^*)  \bigg)
\end{align}
\noindent where $S_{\rm {n}{m}}$ is known as the Debye-Weller factor:
\begin{align}
S_{\rm {n}{m}}= \exp\left[\sum_{\rm {q}} \bigg\lbrace - \left(\frac{|\lambda_{\rm {n}{q}}|^2}{2} + \frac{|\lambda_{\rm {m}{q}}|^2}{2}\right) + \lambda_{\rm {n}{q}}^*\lambda_{\rm {m}{q}}\bigg\rbrace \right]
\label{eq:Snm}
\end{align}
% Defining the Lagrangian in the following way:
%\begin{equation}
%L=\bra{D_1(t)} \frac{i}{2} \left( \overset{\rightarrow}{\frac{\partial}{\partial t}}- \overset{\leftarrow}{\frac{\partial}{\partial t}}\right)-H \ket{D_1(t)}
%\end{equation}\label{eq: Lagrangian}
Given the Lagrangian (\ref{eq: Lagrangian}), the Euler-Lagrange equations can be readily obtained:
\begin{eqnarray}
\frac{d}{dt}\left(\frac{\partial \mathcal{L}}{\partial \dot{\alpha}_{\rm {n}}^*}\right)-\frac{\partial \mathcal{L}}{\partial\alpha_{\textbf{\rm {n}}}^*}&=0\label{eq:EOMS1}\\
\frac{d}{dt}\left(\frac{\partial \mathcal{L}}{\partial \dot{\lambda}_{\rm {n}{q}}^*}\right)-\frac{\partial \mathcal{L}}{\partial\lambda_{\rm {n}{q}}^*}&=0\label{eq:EOMS2}
\end{eqnarray}
resulting in the following equations of motion for the variational parameters:

\begin{equation}
\dot{\alpha_{\rm {n}}} = \frac{i}{\hbar}  \bigg[  T_{\rm {n}} +  \alpha_{\rm {n}} R_{\rm {n}} \bigg]
\label{eq:EOMalpha}
\end{equation}
\begin{equation}
\dot{\lambda_{\rm {n}{q}}}=\frac{i}{\hbar}\bigg[ \frac{\Omega_{\rm {n}{q}}}{\alpha_{\rm {n}}} + \frac{1}{\sqrt{N}} \hbar \omega_{\rm {q}} g_{\rm {n}{q}} e^{-i\rm {q}{n}}  - \hbar \omega_{\rm {q}} \lambda_{\rm {n}{q}}   \bigg]
\label{eq:EOMlambda}
\end{equation}
where:
\begin{equation}
T_{\rm {n}}=-  \sum_{\rm {m}}  \alpha_{\rm {m}} J_{\rm {n}{m}}  S_{\rm {n}{m}}
\label{eq:Tn}
\end{equation}
\begin{align}
\Omega_{\rm {n}{q}}=& - \sum_{\rm {n}^\prime} \alpha_{\rm {n}^\prime} J_{\rm {n}{n}^\prime} \bigg(  \lambda_{\rm {n}^\prime {q}} -\lambda_{\rm {n}{q}}\bigg) S_{\rm {n}{n}^\prime}
\label{eq:Omeganq}
\end{align}
\begin{align}
R_{\rm {n}} &= \operatorname{Re} \sum_{\rm {q}} \bigg( \overline{\omega}_{\rm {n}{q}}+  \frac{1}{\sqrt{N}}  \hbar \omega_{\rm {q}} g_{\rm {nq}} e^{-i\rm {q}{n}} \bigg)\lambda_{\rm {nq}}^*\\\nonumber
& =  \operatorname{Re} \sum_{\rm q} \bigg( - \frac{\Omega_{\rm {n}{q}}}{\alpha_{\rm {n}}}+  \frac{1}{\sqrt{N}}  \hbar \omega_{\rm {q}} g_{\rm {nq}} e^{-i\rm {q}{n}} \bigg)\lambda_{\rm {nq}}^*\nonumber
\end{align}
\begin{equation}
\overline{\omega}_{\rm {n}{q}}=i \hbar\dot{\lambda_{\rm {n}{q}}}  - \hbar \omega_{\rm {q}} \lambda_{\rm {n}{q}}
+   \frac{1}{\sqrt{N}}  \hbar \omega_{\rm {q}} g_{\rm {n}{q}} e^{-i\rm {q}{n}}
\end{equation}

\section{Derivation of 2D Spectra}

To simulate the 2D spectra of the system, we add to Eq. (\ref{eq:Hamiltonian}) an additional Hamiltonian term representing the light-matter interaction:
\begin{equation}\label{H_L}
H_{L}=-\sum_{m=1}^N\left[{\mathbf E}({\mathbf r},t)\cdot\boldsymbol{\mu}_m a_m^{\dagger}
+{\mathbf E}({\mathbf r},t)^*\cdot{\boldsymbol{\mu}}_m^* a_m\right],
\end{equation}
Here, ${\mathbf E}$ is the external electric field
and can be described by:
\begin{eqnarray}
{\textbf{E}}(\textbf{r},t)&=&{\textbf{E}}_1(\textbf{r},t)+{\textbf{E}}_2(\textbf{r},t)+{\textbf{E}}_3(\textbf{r},t)\nonumber\\
{\textbf{E}}_1(\textbf{r},t)&=&\textbf{e}_1E_1(t-\tau_1)e^{i\textbf{k}_1\cdot\textbf{r}-i\omega_1t+i\phi_1}\nonumber\\
{\textbf{E}}_2(\textbf{r},t)&=&\textbf{e}_2E_2(t-\tau_2)e^{i\textbf{k}_2\cdot\textbf{r}-i\omega_2t+i\phi_2}\nonumber\\
{\textbf{E}}_3(\textbf{r},t)&=&\textbf{e}_3E_3(t-\tau_3)e^{i\textbf{k}_3\cdot\textbf{r}-i\omega_3t+i\phi_3},
\end{eqnarray}
where $\textbf{e}_a$, $\textbf{k}_a$, $\omega_a$, $E_a(t)$, and $\phi_a$ ($a=1,2,3$) denote the polarization, wave
vector, carrier frequency, dimensionless envelope, and initial phase. It is common to define the pulse arrival times in the system-field
Hamiltonian (\ref{H_L}) as follows: \begin{equation}
\tau_{1}=-T_w-\tau,\,\,\,\tau_{2}=-T_w,\,\,\,\tau_{3}=0,\label{Ttau}\end{equation}
where $\tau$ (the so-called coherence time) is the delay time between
the second and the first pulse, and $T_w$ (the so-called population
time) is the delay time between the third and the second pulse.
Before the optical excitation ($t\ll-T_w-\tau$) the system is assumed to be in its global ground state $|g\rangle|0\rangle_{\rm ph}$
($|0\rangle_{\rm ph}$ is the vacuum state of the prime phonons), while the heat bath  is in thermal equilibrium at temperature $T_{eq}$.
For the short pulse limit, we have $E_a(t)=E_0\delta(t)$. $\boldsymbol{\mu}_m$ is the dipole matrix element.
The polarization operator $\hat{\textbf{P}}$ is given by:
\begin{equation}
\hat{\textbf{P}}=\sum_{m=1}^N\boldsymbol{\mu}_m(a_m^{\dagger}+a_m),
\end{equation}

The one-exciton response functions are given by:
\begin{eqnarray}\label{response functions}
&&R_1(\tau,T_w,t)=\sum_{mm_1m_2m_3}\boldsymbol{\mu}_{m_2}^*(\textbf{e}_1\cdot\boldsymbol{\mu}_{m_3})(\textbf{e}_2^*\cdot\boldsymbol{\mu}^*_{m})\nonumber\\
&&(\textbf{e}_3\cdot\boldsymbol{\mu}_{m_1})\alpha^*_{m_1m}(T_w)\alpha_{m_2m_3}(\tau+T_w+t)\nonumber\\
&&e^{-\frac 1 2\sum_q(|\lambda_{m_1q}(T_w)|^2+|\lambda_{m_2q}(\tau+T_w+t)|^2)}\nonumber\\
&&e^{\lambda^*_{m_1q}(T_w)\lambda_{m_2q}(\tau+T_w+t)e^{i\omega_qt}}F_1^I(\tau,T_w,t),\nonumber\\
&&R_2(\tau,T_w,t)=\sum_{mm_1m_2m_3}\boldsymbol{\mu}_{m_2}^*(\textbf{e}_1^*\cdot\boldsymbol{\mu}_{m}^*)(\textbf{e}_2\cdot\boldsymbol{\mu}_{m_3})\nonumber\\
&&(\textbf{e}_3\cdot\boldsymbol{\mu}_{m_1})\alpha^*_{m_1m}(\tau+T_w)\alpha_{m_2m_3}(T_w+t)\nonumber\\
&&e^{-\frac 1 2\sum_q(|\lambda_{m_1q}(\tau+T_w)|^2+|\lambda_{m_2q}(T_w+t)|^2)}\nonumber\\
&&e^{\lambda^*_{m_1q}(\tau+T_w)\lambda_{m_2q}(T_w+t)e^{i\omega_qt}}F_2^I(\tau,T_w,t),\nonumber\\
&&R_3(\tau,T_w,t)=\sum_{mm_1m_2m_3}\boldsymbol{\mu}_{m_2}^*(\textbf{e}_1^*\cdot\boldsymbol{\mu}_{m}^*)(\textbf{e}_2\cdot\boldsymbol{\mu}_{m_1})\nonumber\\
&&(\textbf{e}_3\cdot\boldsymbol{\mu}_{m_3})\alpha^*_{m_1m}(\tau)\alpha_{m_2m_3}(t)\nonumber\\
&&e^{-\frac 1 2\sum_q(|\lambda_{m_1q}(\tau)|^2+|\lambda_{m_2q}(t)|^2)}\nonumber\\
&&e^{\lambda^*_{m_1q}(\tau)\lambda_{m_2q}(t)e^{i\omega_q(T_w+t)}}F_3^I(\tau,T_w,t),\nonumber\\
&&R_4(\tau,T_w,t)=\sum_{mm_1m_2m_3}\boldsymbol{\mu}_{m}^*(\textbf{e}_1\cdot\boldsymbol{\mu}_{m_3})(\textbf{e}_2^*\cdot\boldsymbol{\mu}^*_{m_2})\nonumber\\
&&(\textbf{e}_3\cdot\boldsymbol{\mu}_{m_1})\alpha^*_{m_1m}(-t)\alpha_{m_2m_3}(\tau)\nonumber\\
&&e^{-\frac 1 2\sum_q(|\lambda_{m_1q}(-t)|^2+|\lambda_{m_2q}(\tau)|^2)}\nonumber\\
&&e^{\lambda^*_{m_1q}(-t)\lambda_{m_2q}(\tau)e^{-i\omega_qT_w}}F_4^I(\tau,T_w,t),\nonumber\\
\end{eqnarray}
where $\alpha_{m_1m}(t)$ represent the probability amplitude at a time $t$ for the exciton at the site ${\rm m_1}$ with the initial state $|m\rangle$, and $\lambda_{m_1q}(t)$ are the corresponding displacement of phonon. $F_{1-4}^I(\tau,T_w,t)$ are the lineshape factors of the response functions $R_{1-4}$.  The dissipative and dephasing effects of the low-frequency bath are enclosed in the line-broadening function $g(t)$,
\begin{align}
g(t)=\int_{0}^\infty d\omega \frac{J_{\rm{Bath}}(\omega)}{\omega^2} \bigg[ & \coth\frac{\hbar\omega}{2k_B T}(1-\cos\omega t)\\\nonumber
& +i(\sin\omega t - \omega t)\bigg]
\end{align}
\noindent which provides spectral broadening of sharp features introducing a dephasing of the multi-dimensional response functions. We will assume the line-broadening function of the whole aggregate equals the one for a single chromophore by virtue of the global nature of the thermal bath. In this expression, $J_{\rm{Bath}}(\omega)$ is the bath's spectral density, $k_{\rm{B}}$ is the Boltzmann's constant and $T$ is the temperature. Regarding the structure of the bath, we use a brownian oscillator model of the Drude-Lorentz type whose spectral density is the following:
\begin{equation}\label{eq:Drude}
J_{\rm{Bath}}(\omega)=2\lambda_0\frac{\omega\gamma_0}{\omega^2+\gamma_0^2}
\end{equation}
with reorganization energy $\lambda_0$ and relaxation time $\gamma_0$. The corresponding line-broadening function can be readily calculated:
\begin{eqnarray}
&g(t)&=\frac{\lambda_0}{\gamma_0} \cot\left(\frac{\hbar \beta \gamma_0}{2}\right) \left[ e^{-\gamma_0 t} + \gamma_0 t -1\right]\\\nonumber
&+&  \frac{4\lambda_0\gamma_0}{\hbar \beta} \sum_{n=1}^\infty\frac{e^{-\nu_n t}+\nu_n t -1}{\nu_n(\nu_n^2-\gamma_0^2)}  -i \frac{\lambda_0}{\gamma_0}\left[ e^{-\gamma_0 t} + \gamma_0 t -1\right]
\end{eqnarray}
\noindent where $\nu_n=2\pi n/\hbar \beta $ are the so-called Matsubara frequencies and $\beta=1/k_{\rm{B}}T$ is the inverse temperature. Assuming that the system-bath coupling is the same for all excitons, we can obtain the lineshape factors
within the second-order cummulant expansion. They are determined by \cite{MukamelBook}:
\begin{eqnarray}\label{lineshape}
F_1^{I}(\tau,T_w,t)&=&\exp[-g^*(t)-g(\tau)-g^*(T_w)+g^*(T_w+t)\nonumber\\
&&+g(\tau+T_w)-g(\tau+T_w+t)],\nonumber\\
F_2^{I}(\tau,T_w,t)&=&\exp[-g^*(t)-g^*(\tau)+g(T_w)-g(T_w+t)\nonumber\\
&&-g^*(\tau+T_w)+g^*(\tau+T_w+t)],\nonumber\\
F_3^{I}(\tau,T_w,t)&=&\exp[-g(t)-g^*(\tau)+g^*(T_w)-g^*(T_w+t)\nonumber\\
&&-g^*(\tau+T_w)+g^*(\tau+T_w+t)],\nonumber\\
F_4^{I}(\tau,T_w,t)&=&\exp[-g(t)-g(\tau)-g(T_w)+g(T_w+t)\nonumber\\
&&+g(\tau+T_w)-g(\tau+T_w+t)].
\end{eqnarray}
In addition, we can define
\begin{equation}\label{C4}
C_{m,m_1,m_2,m_3}=(\boldsymbol{e}_{1} \boldsymbol{\mu}_{m})(\boldsymbol{e}_{2} \boldsymbol{\mu}_{m_1})(\boldsymbol{e}_{3} \boldsymbol{\mu}_{m_2})(\boldsymbol{e}_{4} \boldsymbol{\mu}_{m_3})
\end{equation}
which are the geometrical factors which must be averaged over the orientations of the transition dipole moments $\boldsymbol{\mu}_{m}$. By assuming, for simplicity, that all laser beams possess the same polarization, we obtain
\begin{eqnarray}\label{C4a}
\bar{C}_{m,m_1,m_2,m_3}&=&\frac{1}{15}((\boldsymbol{\mu}_{m}\boldsymbol{\mu}_{m_1})(\boldsymbol{\mu}_{m_2}\boldsymbol{\mu}_{m_3})\nonumber\\
&&+(\boldsymbol{\mu}_{m}\boldsymbol{\mu}_{m_2})(\boldsymbol{\mu}_{m_1}\boldsymbol{\mu}_{m_3})\\
&&+(\boldsymbol{\mu}_{m}\boldsymbol{\mu}_{m_3})(\boldsymbol{\mu}_{m_2}\boldsymbol{\mu}_{m_1})).\nonumber\\
\end{eqnarray}
In the calculation of Eqs.~(\ref{response functions}), we have used the average coefficients $\bar{C}$.\\

\section{Accuracy of the variational state}
The accuracy of the variational Davydov wave function has been extensively analyzed in the past \cite{Luo2010,Sun2013} and multiple Davydov trial states have been recently proved to provide \cite{Zhou2015,Zhou2016} accurate results comparable with numerically exact methods such as HEOM. The addition of static disorder does not impact the performance of the variational formalism in any noticeable way and the reader is referred to Ref.~\cite{Luo2010,Sun2013} for an extensive survey. Accuracy is tested as a measure of the deviation of the trial wave function from the Schr{\"o}dinger equation:
\begin{equation}
\ket{\delta(t)}=i \frac{d}{d t}\ket{D_1}-H\ket{D_1}
\end{equation}
The norm of the deviation vector is thus helpful to analyse the accuracy of the variational solution:
\begin{equation}
\Delta(t)=\sqrt{\braket{\delta
(t)|\delta(t)}}
\label{eq:devamplitude}
\end{equation}
The accuracy of the \textit{ansatz} is finally measured by the relative deviation, defined by:
\begin{equation}
\sigma \equiv \frac{\max \Delta(t)}{\rm{avg}\, E_{ph}(t)}\quad t\in \left[0, t_{\rm{max}}\right]
\label{eq:reldeviation}
\end{equation}
\noindent where $t_{\rm{max}}$ is the total simulation time. The accuracy of the \textit{ansatz} improves with stronger coupling due to the inability of coherent states to capture the {plane wave-like} behaviour of phonon wave functions in the low coupling regime.
{
\section{Numerical Considerations}
The approach that we follow for quantum dynamics in this work includes explicitly the bath degrees of freedom through the coherent states of the Davydov ansatz. Each site of the LH2 is coupled to a multiplicity of modes, which makes ensemble averages of single trajectories computationally demanding. The main advantage of GPU computing is its efficiency in dealing with high dimensional systems, as we did for the chlorosome in a previous work \cite{Somoza2016}.  The LH2 ring is a relatively small system, when compared to other photosynthetic systems such as the chlorosome. In the present context, we exploit GPU computing to simulate a disordered ensemble consisting of thousands of Hamiltonians running simultaneously in parallel, outperforming CPU calculations where individual trajectories are run in series. Regarding the complexity scaling of the algorithm, we can provide further details: Within the framework of the $\rm D_{1}$ ansatz, the total number of variational parameters in a system of $N$ chromophores is $N^2 + N$. For instance, the simulation of 200 fs dynamics in system of 360 sites (approximately 130 000 equations of motion) is accomplished in 5 hours (including energy dynamics, coherence size and error analysis calculations), while this same calculation in a CPU would take more than one day to be completed.}

\bibliographystyle{h-physrev}

\end{document}